\documentclass[pra, onecolumn, longbibliography, floatfix, superscriptaddress]{revtex4-1}
\usepackage{dcolumn}
\usepackage{bm}
\usepackage{graphicx}
\usepackage{amsmath}
\usepackage{enumitem}
\usepackage{latexsym}
\usepackage{amsfonts}
\usepackage{amssymb}
\usepackage{array}
\usepackage{epsfig}
\usepackage[dvipsnames]{xcolor}

\usepackage{subcaption}

\usepackage{color}
\usepackage[colorlinks=true,linkcolor=blue,urlcolor=blue,citecolor=blue,pdfusetitle]{hyperref}
\usepackage{hyperref}
\usepackage{physics}
\usepackage{bbold}

\begin{document}

\title{Witnessing Objectivity on a Quantum Computer}

\author{Dario A. Chisholm}
\affiliation{Universit$\grave{a}$  degli Studi di Palermo, Dipartimento di Fisica e Chimica - Emilio Segr\`e, via Archirafi 36, I-90123 Palermo, Italy}
\affiliation{QTF Centre of Excellence, Department of Physics, Faculty of Science, University of Helsinki, FI-00014 Helsinki, Finland}
\author{Guillermo Garc\'{i}a-P\'{e}rez}
\affiliation{QTF Centre of Excellence, Department of Physics, Faculty of Science, University of Helsinki, FI-00014 Helsinki, Finland}
\affiliation{Complex Systems Research Group, Department of Mathematics and Statistics, University of Turku, FI-20014 Turun Yliopisto, Finland}
\affiliation{Algorithmiq Ltd, Linnankatu 55 K 329, 20100 Turku, Finland}
\author{Matteo A. C. Rossi}
\affiliation{QTF Centre of Excellence, Center for Quantum Engineering,
Department of Applied Physics, Aalto University School of Science, FIN-00076 Aalto, Finland}
\affiliation{Algorithmiq Ltd, Linnankatu 55 K 329, 20100 Turku, Finland}
\author{Sabrina Maniscalco}
\affiliation{QTF Centre of Excellence, Department of Physics, Faculty of Science, University of Helsinki, FI-00014 Helsinki, Finland}
\affiliation{QTF Centre of Excellence, Center for Quantum Engineering,
Department of Applied Physics, Aalto University School of Science, FIN-00076 Aalto, Finland}
\affiliation{Algorithmiq Ltd, Linnankatu 55 K 329, 20100 Turku, Finland}
\author{G. Massimo Palma}
\affiliation{Universit$\grave{a}$  degli Studi di Palermo, Dipartimento di Fisica e Chimica - Emilio Segr\`e, via Archirafi 36, I-90123 Palermo, Italy}
\affiliation{NEST, Istituto Nanoscienze-CNR, Piazza S. Silvestro 12, 56127 Pisa, Italy}

\begin{abstract}

Understanding the emergence of objectivity from the quantum realm has been a long standing issue strongly related to the quantum to classical crossover. Quantum Darwinism provides an answer, interpreting objectivity as consensus between independent observers.
Quantum computers provide an interesting platform for such experimental investigation of quantum Darwinism, fulfilling their initial intended purpose as quantum simulators.
Here we assess to what degree current NISQ devices can be used as experimental platforms in the field of quantum Darwinism.
We do this by simulating an exactly solvable stochastic collision model, taking advantage of the analytical solution to benchmark the experimental results.
	
\end{abstract}
\date{\today}
\maketitle
\section{Introduction}

The concept of objectivity (or more precisely inter-subjectivity) plays a central role in open questions such as the quantum to classical crossover and the measurement problem~\cite{RevModPhys.75.715, PhysRevD.24.1516}. It is thus of great interest to be able to identify and quantify objectivity for a quantum state.

One of the most successful theories in explaining the emergence of objectivity is quantum Darwinism (QD)~\cite{Zurek2009, PhysRevA.73.062310}.
The core idea of QD is that as the system decoheres due to the interaction with an environment~\cite{breuer2002theory, Rivas_2014}, information about the system is encoded into said environment. If the information encoding is redundant, meaning that even a small fraction of the environment has a significant amount of information about the system, and that enlarging the environmental fraction does not offer any additional information, then several different observers can each access a fraction of the environment and independently infer information about the system. In this case they can reach a consensus, and if several observers agree on the state of the system then we can claim that the state is objective.

While a quantum state can never be considered objective \textit{in itself}, objectivity may emerge from the quantum state being part of a larger context~\cite{Auffeves2016}.
Objectivity then becomes a property of the system, dependent on the joint properties of the system and its context, in other words, the environment that caused the decoherence process in the first place.

In QD the signature of objectivity is given by the quantum mutual information (QMI) between the system $\mathcal{S}$ and an environmental fraction $f\mathcal{E}$ defined as $\mathcal{I}(\mathcal{S}:f\mathcal{E})=H(\mathcal{S})+H(f\mathcal{E})-H(\mathcal{S}f\mathcal{E})$, here $H(\cdot)$ is the von Neumann entropy defined as $H(\rho)=-\mathrm{tr}(\rho\ln\rho)$.

In objective states, the redundancy of information is made manifest by the behaviour of $\mathcal{I}(\mathcal{S}:f\mathcal{E})$ as a function of the environmental fraction size $f$, which presents a characteristic plateau that is due to the redundant information encoding. On the other hand $\mathcal{I}(\mathcal{S}:f\mathcal{E})$ for states randomly extracted from the Hilbert space has a distinctively different shape, characterised by almost no QMI between the system and the environmental fraction, if not when the fraction size $f$ is almost half of the total environment.
Let us remark that $\mathcal{I}(\mathcal{S}:f\mathcal{E})$ is the mutual information between the system and a specific environmental fraction of size $f$, and may depend on the chosen fraction; in order to evaluate QD we must in general do the average of the mutual information with all possible environmental fractions of the same size~\cite{Blume-Kohout2005}, we will refer to this quantity as $\mathcal{\Tilde{I}}(\mathcal{S}:f\mathcal{E})$.

\begin{figure}
    \centering
    \includegraphics[trim={0 0cm 0 0cm}, clip, width=0.7\textwidth]{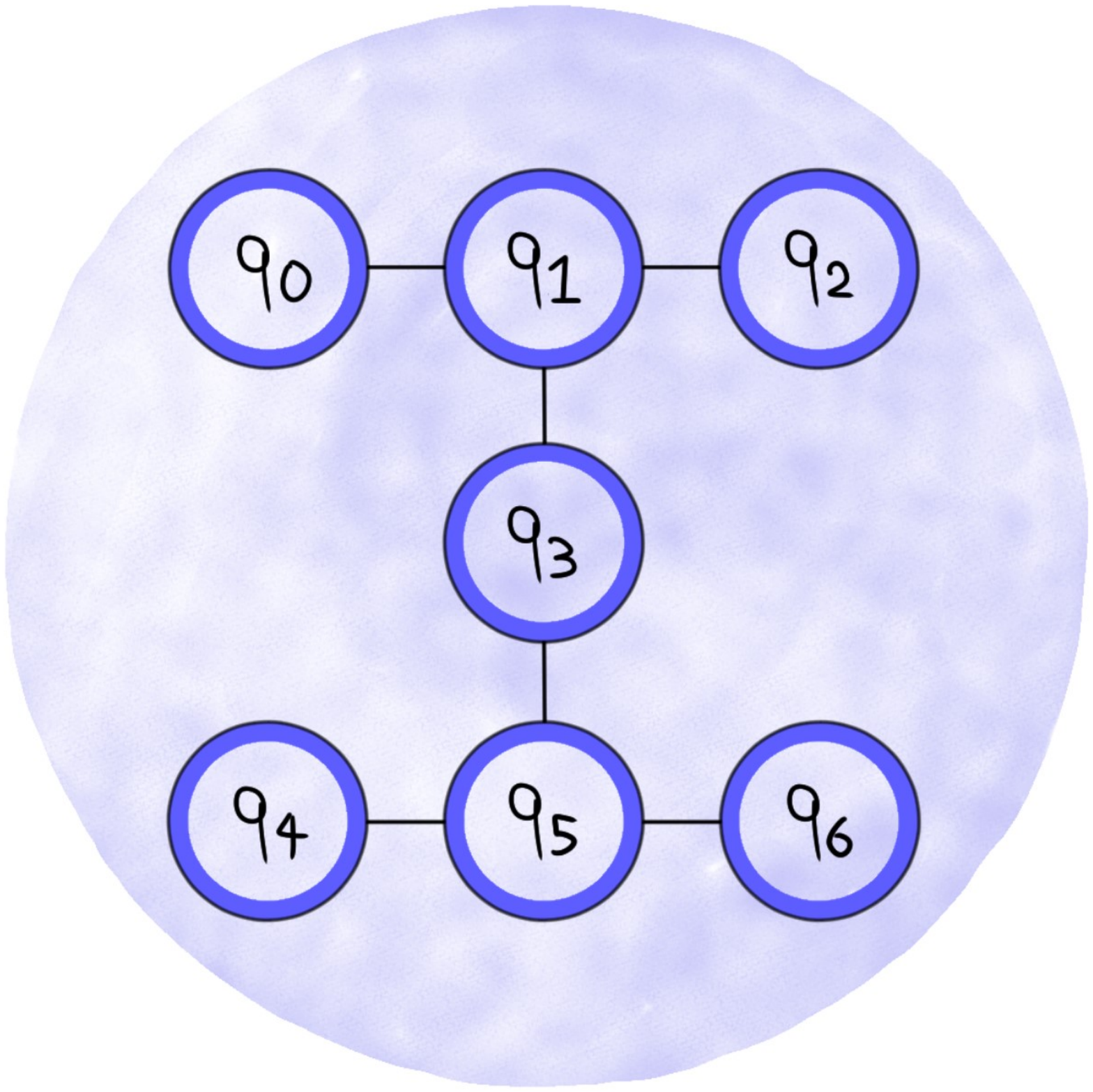}
    \caption{\textbf{Topology of the \texttt{ibmq\_casablanca} quantum computer}. Each blue circle represents a qubit, the fact that two qubits are connected by a line means that it is possible to apply two-qubit gates between them. The symbols ranging from $q_{0}$ to $q_{6}$ are the labels by which the qubits are referred to.}
    \label{fig:topo}
\end{figure}

As it is necessary to keep track not only of the system but also of the environment, experimental investigation of QD is particularly challenging, and so far has been witnessed in the case where the environment is made of three~\cite{PhysRevA.98.020101}, four~\cite{PhysRevLett.123.140402} and five~\cite{CHEN2019580} qubits, in all cases by using optical setups, or through quantum optical interrogation schemes. On the other hand, there has been extensive theoretical study of QD in recent years~\cite{Jess_Riedel_2012, PhysRevA.92.022105, PhysRevA.96.012120, PhysRevA.96.062105, PhysRevA.99.042103, PhysRevA.100.012101, PhysRevA.100.052110, PhysRevResearch.2.013164, Le_2020, e23080995, touil2021eavesdropping, RYAN2021127675}.

From the very beginning of their conception, quantum computers were thought as simulators for physical quantum systems~\cite{Feynman1982, doi:10.1126/science.273.5278.1073}, considering that the very structure of the quantum theory makes simulations of quantum systems using classical resources particularly challenging.

In recent years, the development of Noisy Intermediate-Scale Quantum (NISQ) devices reached a level that makes their usage as quantum simulators possible. This has been shown in several topics including chemistry~\cite{Peruzzo2014, Kandala2017}, open quantum systems~\cite{Garcia-Perez2020} and quantum machine learning~\cite{abbas2020power}. With the fast technological advancement it is reasonable to assume that NISQ devices will soon enable quantum simulations currently unattainable with classical resources alone. The purpose of this work is to investigate their potential to simulate quantum dynamics leading to the emergence of objectivity in the context of QD.

For this purpose we simulate an exactly solvable stochastic collision model (SCM)~\cite{ciccarello2021quantum} first introduced in~\cite{PhysRevResearch.2.012061}.
The fact that it is an exactly solvable model makes it an ideal use case to benchmark the possibility to realise objective states in a quantum computer.
The model is particularly rich because it can exhibit QD and non-Markovianity~\cite{Rivas_2014}, and the system encodes information into the environment in a strictly non-local way.

\section{The model}

The studied model is a stochastic collision model (SCM), a particular class of collision models where the time intervals between system-environment interactions are extracted from a probability distribution.
SCMs where first introduced in~\cite{PhysRevResearch.2.012061} and further explored in~\cite{Chisholm_2021}, see~\cite{doi:10.1142/S0219749914610115, PhysRevLett.117.230401, Vacchini2020} for closely related works.

Similar to collision models, they can describe a wide variety of behaviours by appropriately choosing the initial states of the ancillae and the interaction Hamiltonian between system and ancilla.
As we are interested in a pure dephasing model, we assume for the ancillae to be qubits initially in the $\ket{0}$ state and that the collision between the system $\mathcal{S}$ and an ancilla $\mathcal{A}$ is driven by the Hamiltonian $H_{I}=\frac{\eta}{2}\sigma^{x}_{\mathcal{A}}\otimes\sigma^{z}_{\mathcal{S}}$. Each ancilla interacts only once with the system. The time at which the interaction takes place is randomly distributed according to a Poisson process with rate $\lambda$. Therefore, at any given time $t$, each ancilla has a probability $p(t)=1-e^{-\lambda t}$ to have already interacted with the system.
For conceptual simplicity we assume that collisions are almost instantaneous and define the collision strength as $\theta=\lim_{\tau\rightarrow0}\tau\eta$, with $\tau$ the duration of the collision.

The effect of a single collision between an ancilla and the system is given by the map $\Phi$ that can be expressed in terms of the Kraus operators $K$ and $K^{\dagger}$
\begin{equation*}
\Phi(\rho_{\mathcal{S}})=\frac{1}{2}\bigg[K\rho K^{\dagger}+
K^{\dagger}\rho K\bigg],
\end{equation*}
with
\begin{equation*}
K=
\begin{pmatrix}
		e^{-i\frac{\theta}{2}}&0\\
		0&e^{i\frac{\theta}{2}}\\
\end{pmatrix}.
\end{equation*}
The parameter $\theta$ effectively governs the amount of entanglement being produced after a single collision, and for $\theta=\pi$ there is no entanglement production whatsoever between the system and the ancillae.
Since the collisions are probabilistic, at any given time the total number of collisions is a random variable with a well-defined probability distribution. The state of the system (and the environment as well) is then the average over all possible stochastic realizations of the collision dynamics.

In a SCM the total number af ancillae (and thus the total number of possible collisions) can either be infinite or finite. If the number of ancillae is infinite, the dynamics of the state of the system $\rho_{\mathcal{S}}$ is described by the Lindblad master equation~\cite{Lindblad1976, doi:10.1063/1.522979}
\begin{equation}
	\frac{d\rho_{\mathcal{S}}}{dt}=
	-i[H_{\mathcal{S}}, \rho_{\mathcal{S}}(t)]+\lambda\big(\Phi(\rho_{\mathcal{S}}(t))-\rho_{\mathcal{S}}(t)\big)
	\label{eq:ME}
\end{equation}
where $H_{\mathcal{S}}$ is the Hamiltonian of the system. Equation \ref{eq:ME} is clearly a dephasing dynamics with coherence factor $c(t)=\exp[-\lambda(1-\cos \theta)t]$. It is worth stressing that, if $\theta=\pi$, in each single stochastic realization the state of the system is always pure. Also note that the state dynamics only depends on the product $\lambda(1-\cos \theta)$, so that the system is agnostic to the degree of entanglement being produced by the collisions, as long as the collisional rate is scaled appropriately.

If the number of ancillae is finite, at long enough times the finite size effects become relevant, resulting in non-Markovian behaviour and backflow of information according to the BLP definition~\cite{PhysRevLett.103.210401}.
The system dynamics can not thus be described by a Lindblad master equation, but it is still possible to compute the dynamical map, resulting in the coherence factor $c(t)=\bigg[1+(\cos \theta-1)\Big(1-e^{-\lambda t/n}\Big)\bigg]^{n}$. In such a scenario, the system is no longer agnostic to the degree of entanglement production, but it will still be so in the limit of short times.

The non-Markovian behaviour becomes apparent if we consider that for a pure dephasing dynamics, the trace distance between two different initial states with equal populations is proportional to the coherence factor, so that non-monotonicity in the latter implies non-monotonicity in the former and non-Markovianity according to the BLP definition. 

The aim of this work is to simulate this collision model using the IBM Quantum devices, and to recover a signature of non-Markovianity and QD. We are particularly interested in assessing up to what degree current quantum computers can be used as a platform to witness objectivity.

\section{Simulation on a NISQ device}

We study the case where the system, a single qubit, interacts with an environment made of ancillae, each a qubit, according to a SCM. Each ancilla is initially in the $\ket{0}$ state and collides with the system at a random moment in time, such that at any time $t$ there is the probability $p=1-e^{-\lambda t}$ for the ancilla to have already collided with the system. In all simulations we set $\lambda=1$, there is thus a unique correspondence between the physical time $t$ and the probability $p$, we will use this to control the value of the physical time by changing the collision probability.

Given the $\sigma^{x}_{\mathcal{A}}\otimes\sigma^{z}_{\mathcal{S}}$ form of the interaction Hamiltonian, the SCM is a model of pure dephasing in the computational basis. Because of this, the most interesting scenario is the one where the system is initially in an equal weight superposition of the pointer basis (the computational one). We will thus focus on the case where the initial state of the system is $\ket{+}_{\mathcal{S}}=\frac{1}{\sqrt{2}}(\ket{0}_{\mathcal{S}}+\ket{1}_{\mathcal{S}})$.

We will only consider the case of a finite environment, as it is necessary in order to study QD.

The studied model is stochastic in nature, thus it can not be straightforwardly simulated in a quantum computer, which only simulates unitary dynamics. This issue can be overcome by performing a purification of the global state. By introducing an appropriate super-environment as the cause behind the random collision times, one obtains a system-environment-superenvironment model whose resulting dynamic is overall unitary.

All the experiments presented here are performed using \texttt{ibmq\_casablanca}, which is one of the IBM Quantum Falcon Processors with 7 qubits, schematically represented in Fig.~\ref{fig:topo}.

\begin{figure}
    \centering
    
    \subfloat[]
    {\includegraphics[trim={0 0 0 0}, clip, width=0.45\textwidth]{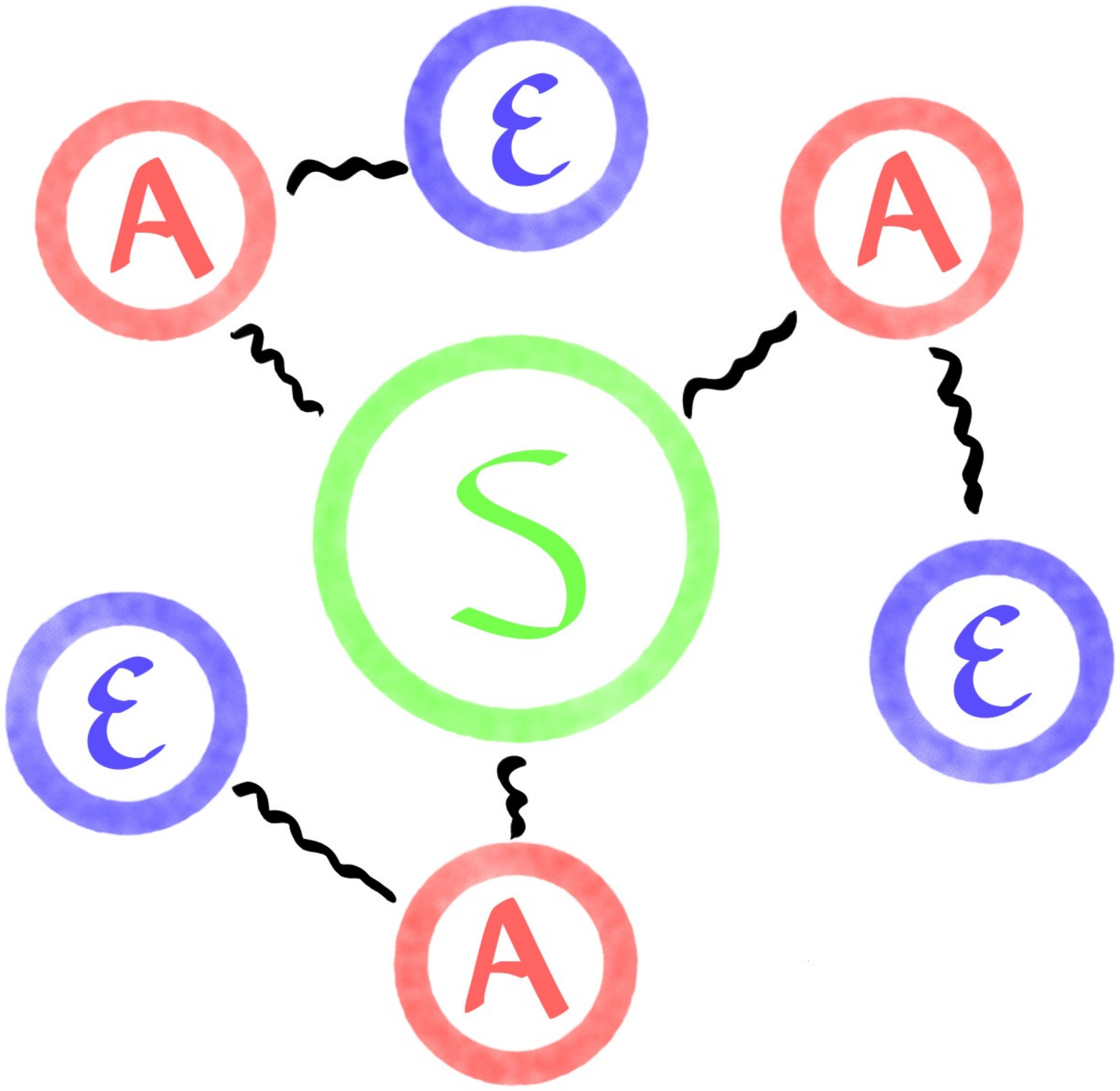}} \quad
    \subfloat[]
    {\includegraphics[trim={0 2cm 0 0}, clip, width=0.5\textwidth]{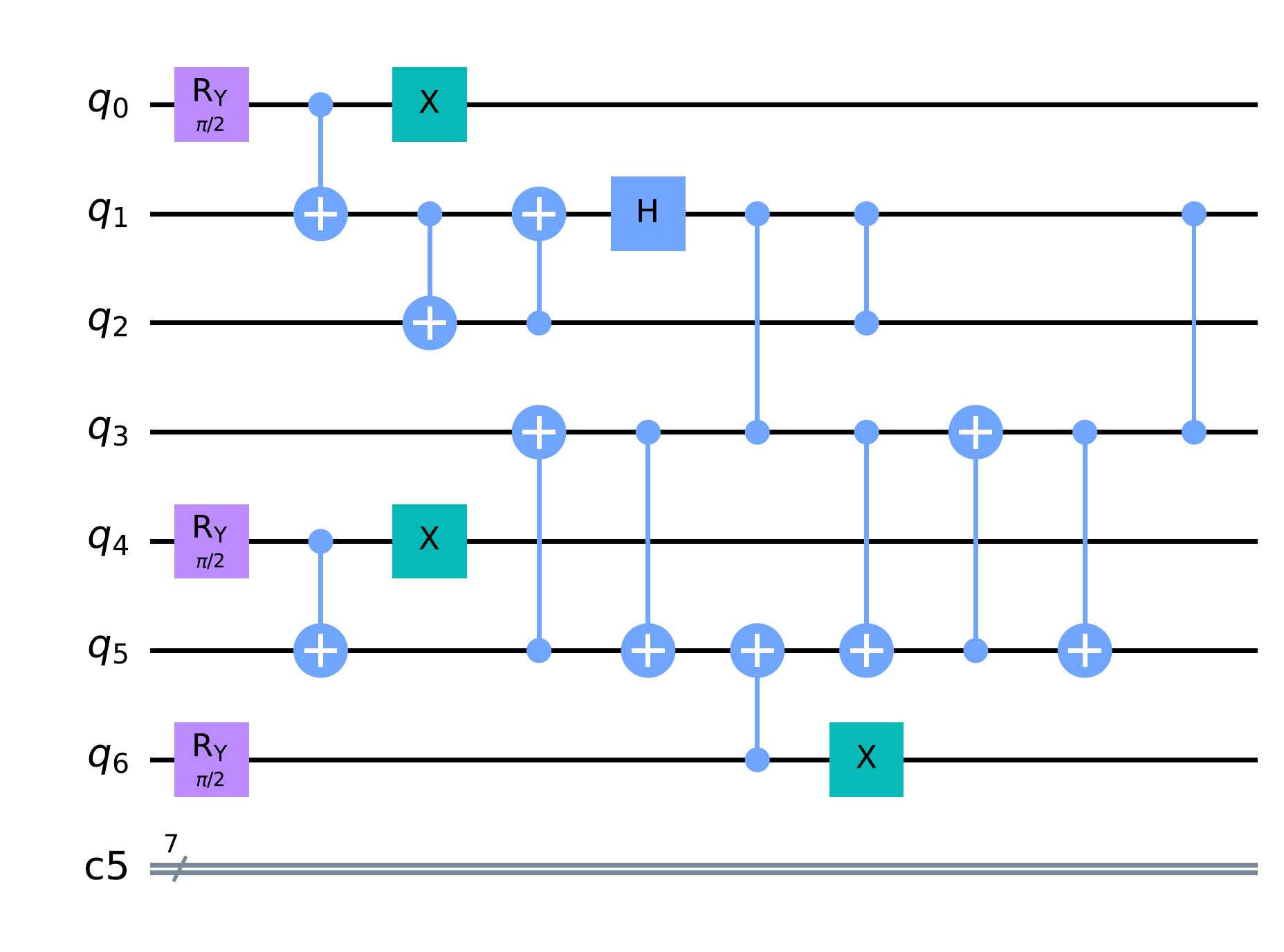}}
    \caption{(a) Visual representation of the process: $\mathcal{S}$ represents the system, $\mathcal{A}$ the ancillae and $\mathcal{E}$ the emitters, the black lines represent the interaction between the different qubits. (b) Circuit representing the gate decomposition of the process: each line represents one of the qubits of the quantum computer. The ancilla-emitter states are prepared with a rotation along the Y axis followed by a CNOT and a local NOT gate. The interaction between the system and each ancilla is a Z gate on the ancilla controlled by the state of the system. The circuit uses several SWAP gates (represented by two or three consecutive CNOTs) in order to transfer the states of the qubits in the appropriate positions. At the end of the circuit, q1 is the system qubit; q2, q3 and q5 are the ancilla qubits; q0, q4 and q6 are the emitter qubits}
    \label{fig:full_cir}
\end{figure}
\subsection{Purification}

The constituents of the super-environment take the name of the emitters of the ancillae. Each ancilla is in an entangled state with its emitter, in a superposition between being un-emitted and being emitted. The weight of the superposition reflects the probability of the ancilla being emitted (and having collided) at any given time.
We assume that each ancilla is emitted in the $\ket{0}_{\mathcal{A}}$ state, and once emitted it immediately collides with the system. For a single emitter-ancilla pair, the global state would then be the following:
\begin{equation}
\begin{split}
\ket{\psi(t)}=&\sqrt{1-p(t)}\ket{1}_{\mathcal{E}}\ket{0}_{\mathcal{A}}\ket{+}_{\mathcal{S}}\\&+
i\sqrt{p(t)}\big(\cos{\theta/2}\ket{0}_{\mathcal{E}}\ket{0}_{\mathcal{A}}\ket{+}_{\mathcal{S}}-i\sin{\theta/2}\ket{0}_{\mathcal{E}}\ket{1}_{\mathcal{A}}\ket{-}_{\mathcal{S}}\big),
\end{split}
\label{eq:state}
\end{equation}
with $p(t)=1-e^{-\lambda t}$. The vectors $\ket{0}_{\mathcal{E}}$ and $\ket{1}_{\mathcal{E}}$ represent the ground and excited states of the emitter, respectively.
Each ancilla thus requires the use of two qubits in the simulation: one for the ancilla itself and one for its emitter.
We will focus on studying the case of non-entangling collisions, as it represents one of the most interesting scenarios described by the model. Moreover, the non-entangling scenario is the only one that can be simulated without system-ancilla unitaries controlled by the emitter, three-qubit gates that would make the simulation too noisy.
\begin{figure}
    \centering
    \includegraphics[width=0.5\textwidth]{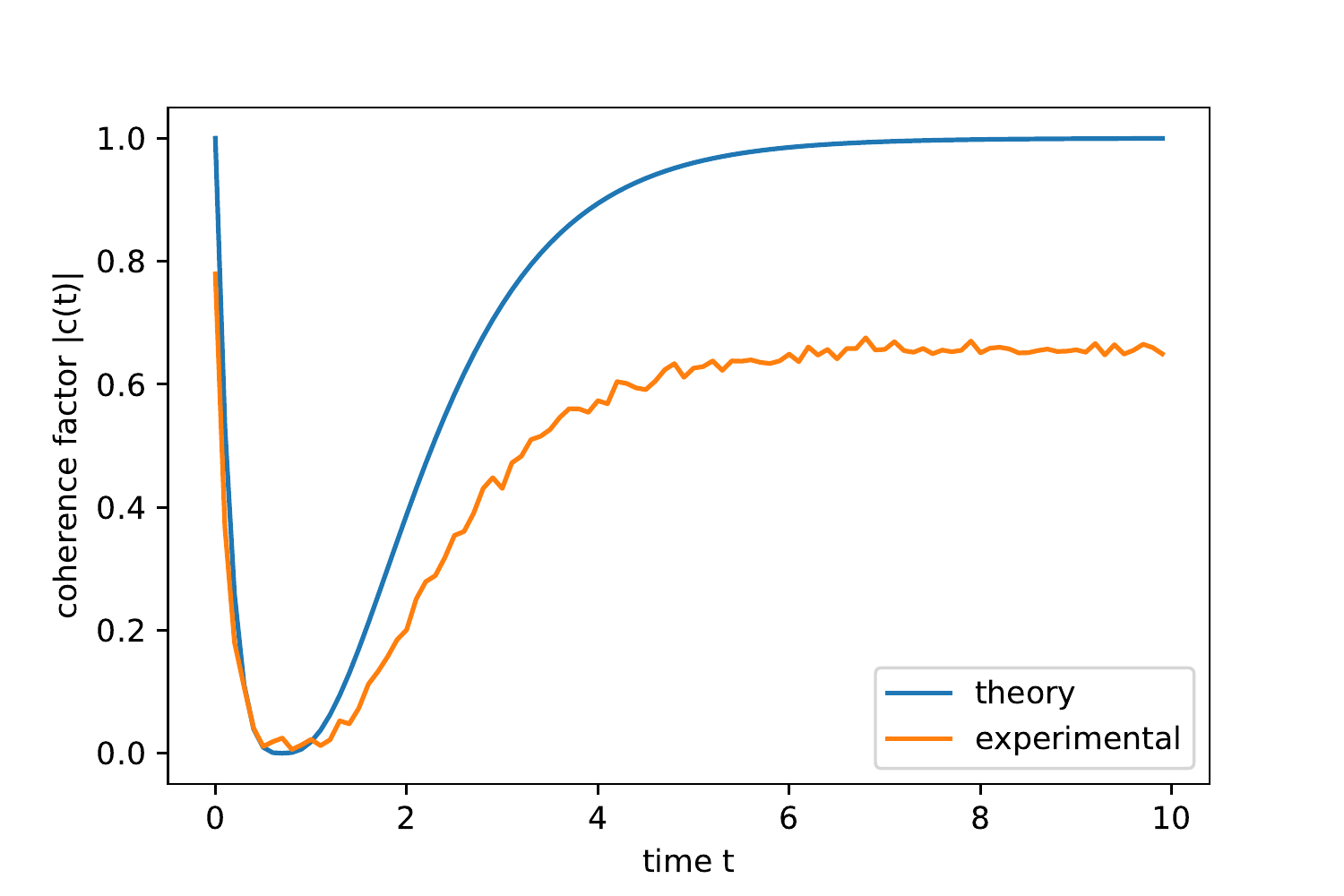}
    \caption{\textbf{coherence factor for the system as a function of time.} Results obtained in the case of three ancillae and three emitters, each represented by an individual qubit. The coherence factor is obtained by performing state tomography at each time step. In this model the non-monotonic behaviour of the coherence factor implies a non-monotonic behaviour of the trace distance between two initially distinguishable states, and can therefore be seen as a signature of non-Markovianity.}
    \label{fig:full_sys_pur}
\end{figure}
We are thus able to simulate a stochastic collision model with three ancillae and three emitters, whose resulting circuit is shown in Fig.~\ref{fig:full_cir}.

At the start of the simulations all qubits are in the $\ket{0}$ state. It is then necessary to prepare the states of ancillae, emitters and system before letting them interact.
The state of each ancilla-emitter pair is prepared by applying a $R_{y}(2\alpha)$ gate on the emitter followed by a CNOT gate on the ancilla controlled by the state of the emitter and finally a NOT gate on the emitter, this results in states of the form   $\sqrt{1-p}\ket{1}_{\mathcal{E}}\ket{0}_{\mathcal{A}}+\sqrt{p}\ket{0}_{\mathcal{E}}\ket{1}_{\mathcal{A}}$.
The NOT gate corresponds to the $\sigma^{x}$ operator. The CNOT is a NOT gate if the state of the control is $\ket{1}$, identity if it is $\ket{0}$, so it takes the form of $\ket{0}\bra{0}\otimes \mathbb{1}+\ket{1}\bra{1}\otimes \sigma^{x}$.
The gate $R_{y}(2\alpha)=e^{-i\alpha\sigma^{y}/2}$ is a rotation around the Y axis of the Bloch sphere. The angle $\alpha$ is the parameter through which we can control the physical time and is given by $\alpha=\arccos(e^{-\lambda t/2})$, so that the probability amplitude for the state in which the ancilla has not been emitted is $\sqrt{1-e^{-\lambda t}}$. 
The state of the system is prepared using a Hadamard gate, resulting in a $\ket{+}_{\mathcal{S}}$ state.
The interaction between system and environment is a Z gate applied on the ancilla controlled by the state of the system, and has the form $\ket{0}\bra{0}\otimes \mathbb{1}+\ket{1}\bra{1}\otimes \sigma^{z}$.

In the \texttt{ibmq\_casablanca} quantum computer, it is not possible for all qubits to interact with one another and the list of possible two-qubit interactions defines the topology of the quantum computer, shown in Fig.~\ref{fig:topo}. To implement all the desired two-qubit gates it is thus necessary to use SWAP gates, in order to transfer the states of the qubits in the appropriate positions. The choice of qubit $q1$ as the system qubit is motivated as being the one that allowed us, given the specific topology of the \texttt{ibmq\_casablanca} quantum computer, to minimise the overall number of SWAP gates. 

\begin{figure}
    \centering
    \includegraphics[width=0.5\textwidth]{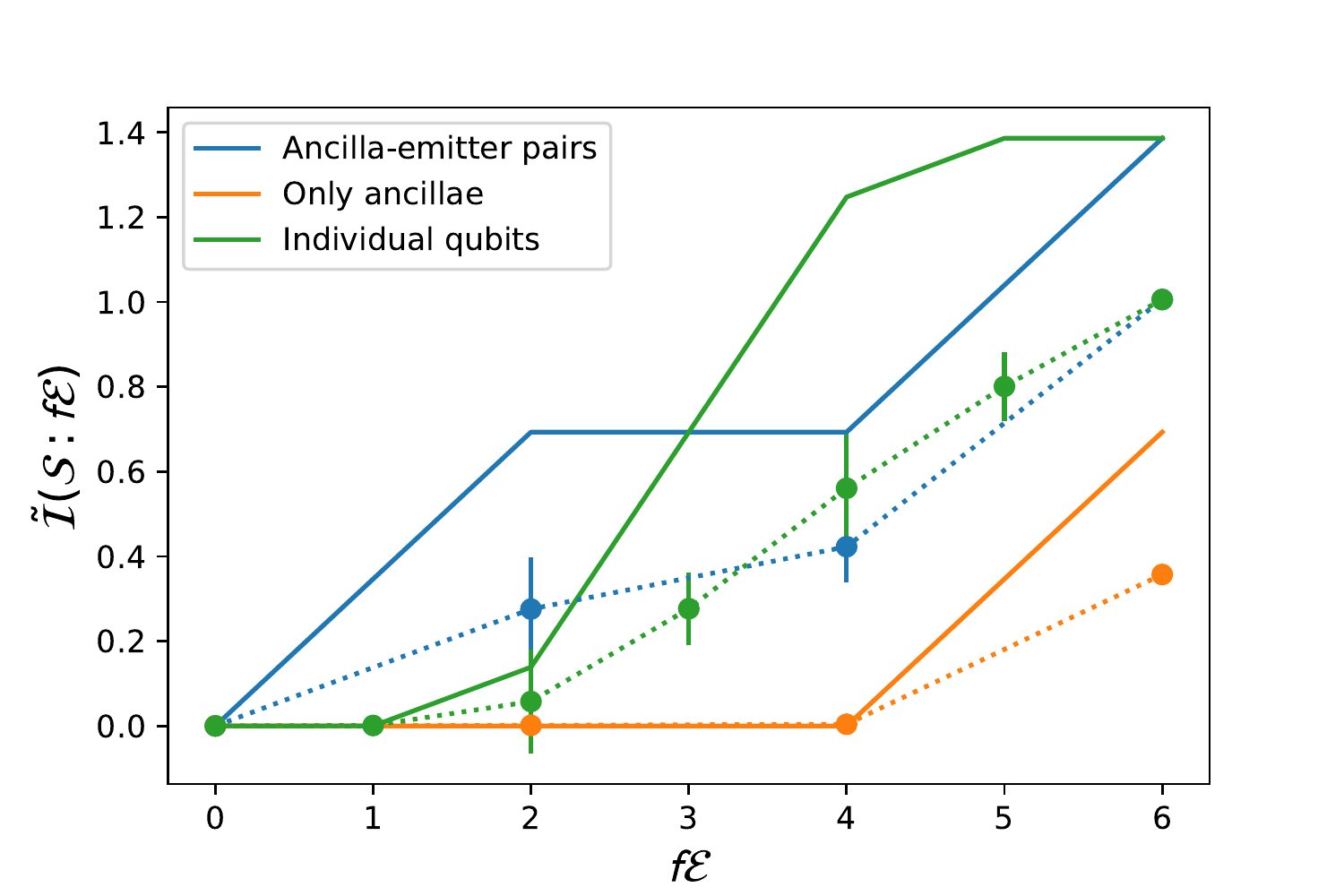}
    \caption{\textbf{Averaged mutual information between system and environmental fraction as a function of the size of the environmental fraction, for an environment of three ancilla-emitter pairs.} The dotted lines represent experimental values, while the continuous lines represent the theoretical expectations.
    The error bars are the standard error of the mean of the mutual information.
    Notice how the experimental values are consistently lower than the theoretical expectations: this is due to the decoherence the quantum computer is subject to, which erodes the mutual information between its components. In the blue line the environment is partitioned into ancilla-emitter pairs, in the orange line the emitters are traced out and only the ancillae are taken into account, in the green line the environemnt is partitioned into the individual environmental qubits. The structure of the partitioning of the environment plays an important role in the emergence of QD.}
    \label{fig:full_dar}
\end{figure}
\begin{figure}
    \centering
    \includegraphics[width=0.5\textwidth]{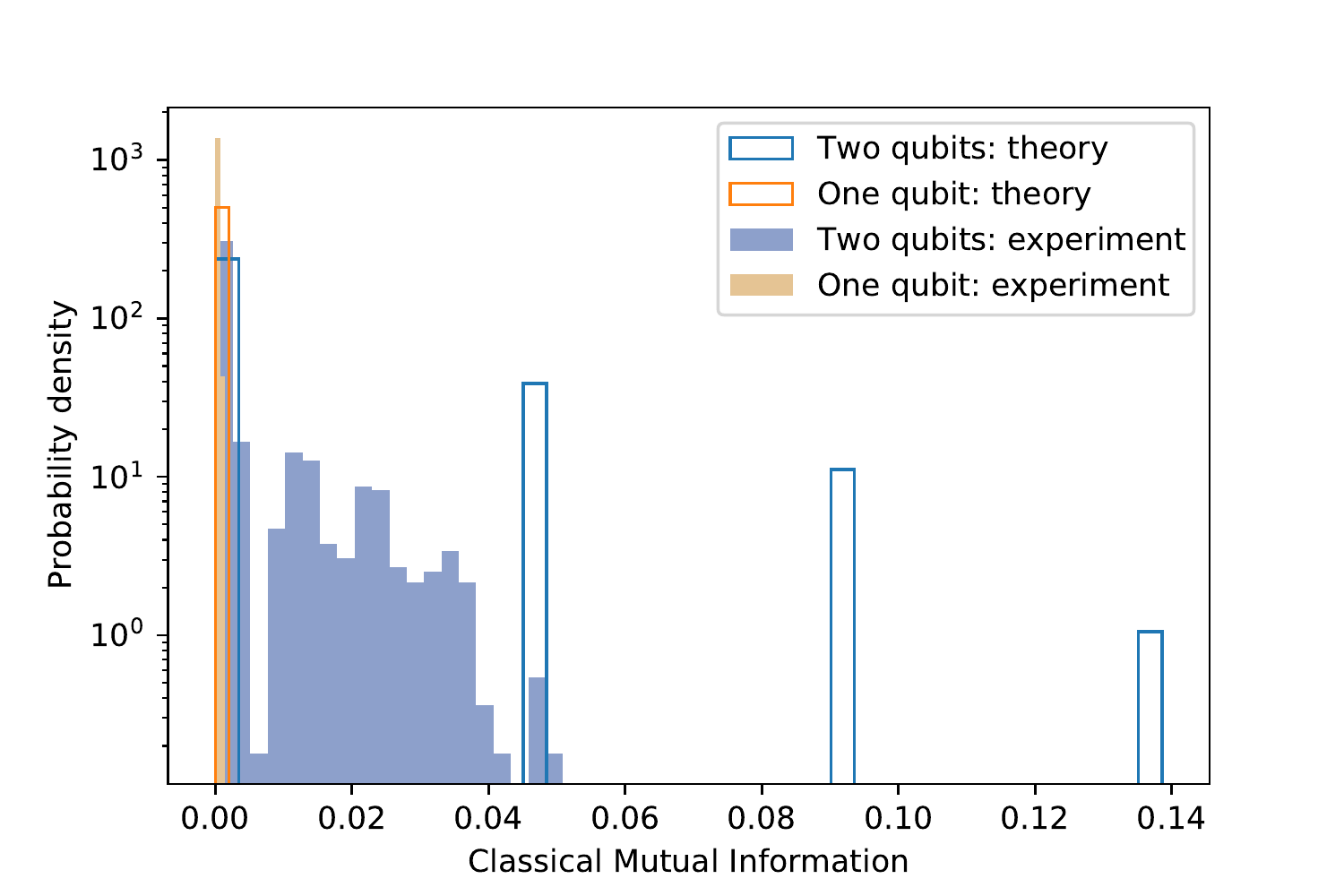}
    \caption{\textbf{Comparison of system-environment correlations between a one-qubit and a two-qubit environment.} 
    In this histogram we represent statistical correlations (classical mutual information) between measurement outcomes of the system and a one-qubit environment (orange), and between measurements of the system and a two-qubit environment (blue). Values are always averaged between all combinations of one-qubit and two-qubit environmental fractions. Measurements are performed in all possible combination of the Pauli basis. Notice how one single qubit holds no correlations with the system: it is necessary to consider at least two qubits simultaneously to witness some correlations with the system.}
    \label{fig:corre}
\end{figure}

The SWAP gate between two generic qubits q1 and q2 is usually obtained by applying three consecutive CNOT gates of the type CNOT(q1, q2) $\rightarrow$ CNOT(q2, q1) $\rightarrow$ CNOT(q1, q2). However, if the state of one of the two qubits is $\ket{0}$, one of the CNOTs is superfluous (an easy way to visualize this is to have the first CNOT controlled by the $\ket{0}$ qubit, which simply results in the identity gate). This helps in reducing the overall number of gates used in the circuit. 

Since the SCM with a finite number of ancillae corresponds to a non-Markovian pure dephasing model, the system undergoes decoherence and recoherence, we show this by running the circuit for different values of the parameter $\alpha$, corresponding to different values of the physical time. For each run of the circuit we perform state tomography of the system qubit, in order to experimentally recover the coherence factor of the system as a function of time, shown in Fig~\ref{fig:full_sys_pur}. Since this is a pure dephasing model, non-monotonicity of the coherence factor implies non-monotonicity in the trace distance between any pair of distinguishable states, and it can thus be seen as a genuine signature of non-Markovianity.

In order to obtain a witness of QD, we obtain the global state by performing a full-state diluted maximum-likelihood tomographic reconstruction~\cite{PhysRevA.75.042108}. We recover the state at the physical time $t=t_{\mathrm{max}}$ at which, according to the theory, the system should be in a maximally mixed state, and there should be objectivity according to QD.

From the global state it is possible to compute the QMI between the system $\mathcal{S}$ and a fraction of the environment $f\mathcal{E}$ using the formula $\mathcal{I}(\mathcal{S}:f\mathcal{E})=H(\mathcal{S})+H(f\mathcal{E})-H(\mathcal{S}f\mathcal{E})$. We stress again that in order to witness objectivity we must evaluate the averaged mutual information $\Tilde{\mathcal{I}}(\mathcal{S}:f\mathcal{E})$, obtained by averaging over all environmental fractions of the same size, more specifically $\Tilde{\mathcal{I}}(\mathcal{S}:f\mathcal{E})=\frac{1}{N}\sum_{\{f\mathcal{E}_{i}\}}\mathcal{I}(\mathcal{S}:f\mathcal{E}_{i})$ where $\{f\mathcal{E}_{i}\}$ is the collection of environmental fractions of size $f$ and $N$ is the number of fractions of size $f$. We show in Fig.~\ref{fig:full_dar} $\Tilde{\mathcal{I}}(\mathcal{S}:f\mathcal{E})$ as a function of the environmental size $f$.

Whether the resulting state is objective or not according to QD strongly depends on the partitioning of the environment. When the individual qubits are taken as the environmental fractions, the mutual information does not exhibit a plateau and objectivity does not seem to emerge. When, however, the environment is partitioned into ancilla-emitter pairs, the resulting mutual information plot is much closer to the one of an objective state. It thus seems to be necessary, for this model, to add an extra degree of information (the partitioning structure) to witness objectivity.

Finally, by tracing out the emitters we recover the mutual information between the system and only the ancillae, where as expected the mutual information is zero unless all three ancillae are simultaneously taken into account.

Having the global state also allows us to address the following point:
The non-Markovian features are present even in the absence of pairwise entanglement between the system and the environmental qubits.
This is manifest not only by concurrence being zero (not shown), but also by the complete absence in statistical correlations in the measurement outcomes of the system and one environmental qubit in all the tomographic measuring bases. It is necessary to take into account at least two environmental qubits at the same time in order to witness any statstical correlations, as shown in Fig~\ref{fig:corre}, where we show the classical mutual information (CMI) between the system and one (two) environmental qubit(s) for all tomographic measurement settings. The CMI is the mutual information of the outcome probability distribution resulting from a measurement operation, and it thus depends on the basis in which measurements are performed.

We can therefore conclude that none of the six environmental qubits has any information of the system when taken individually. It is only when taking account of them together that information about the system manifests, which means that for this model information about the system is encoded in a strictly non-local way.

While it may seem highly artificial that it is necessary to consider the emitters and the ancillae together in order to witness QD, leading to the belief that there is no 'natural' objectivity in this model, this is strongly dependant on the physical model one has in mind. A physical system may be intrinsically partitioned with this structure, allowing a completely ignorant observer to witness objectivity. The opposite may be true as well: if the ancillae are photons emitted by some atoms, it may be physically impossible to measure the ancillae and the emitters together, let alone with the appropriate partition, so that witnessing objectivity would be impossible. As it happens, objectivity (more precisely inter-subjectivity) is a matter of perspective.

\subsection{Condensation protocol}
When collisions are non-entangling, $\theta=\pi$ and $\cos{(\theta/2)}=0$, from Eq.~\ref{eq:state} we can see how the states assumed by the emitter-ancilla pair are spanned by $\ket{1}_{\mathcal{E}}\ket{0}_{\mathcal{A}}$ and $\ket{0}_{\mathcal{E}}\ket{1}_{\mathcal{A}}$. Any physical system that can live in only superpositions of two states is effectively a qubit, so it is possible to encode the information represented by the pair into a single qubit, remapping the previous states into the new $\ket{0}_{\mathcal{C}}$ and $\ket{1}_{\mathcal{C}}$ states. We will call this the \emph{condensed} scenario.

\begin{figure}
    \centering
    \subfloat[]
    {\includegraphics[width=0.45\textwidth]{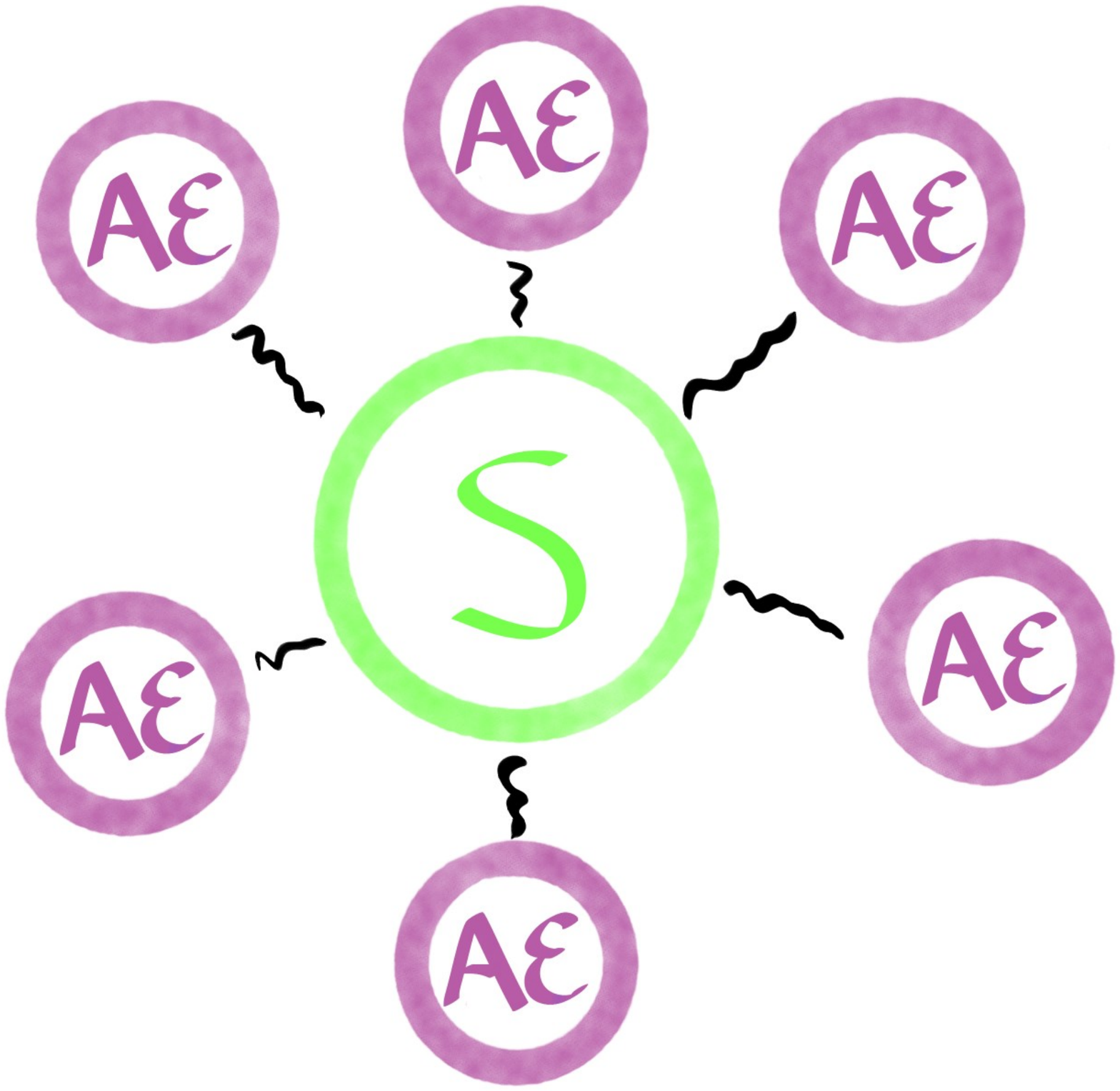}}
    \subfloat[]
    {\includegraphics[trim={0 2cm 0 0}, clip,width=0.5\textwidth]{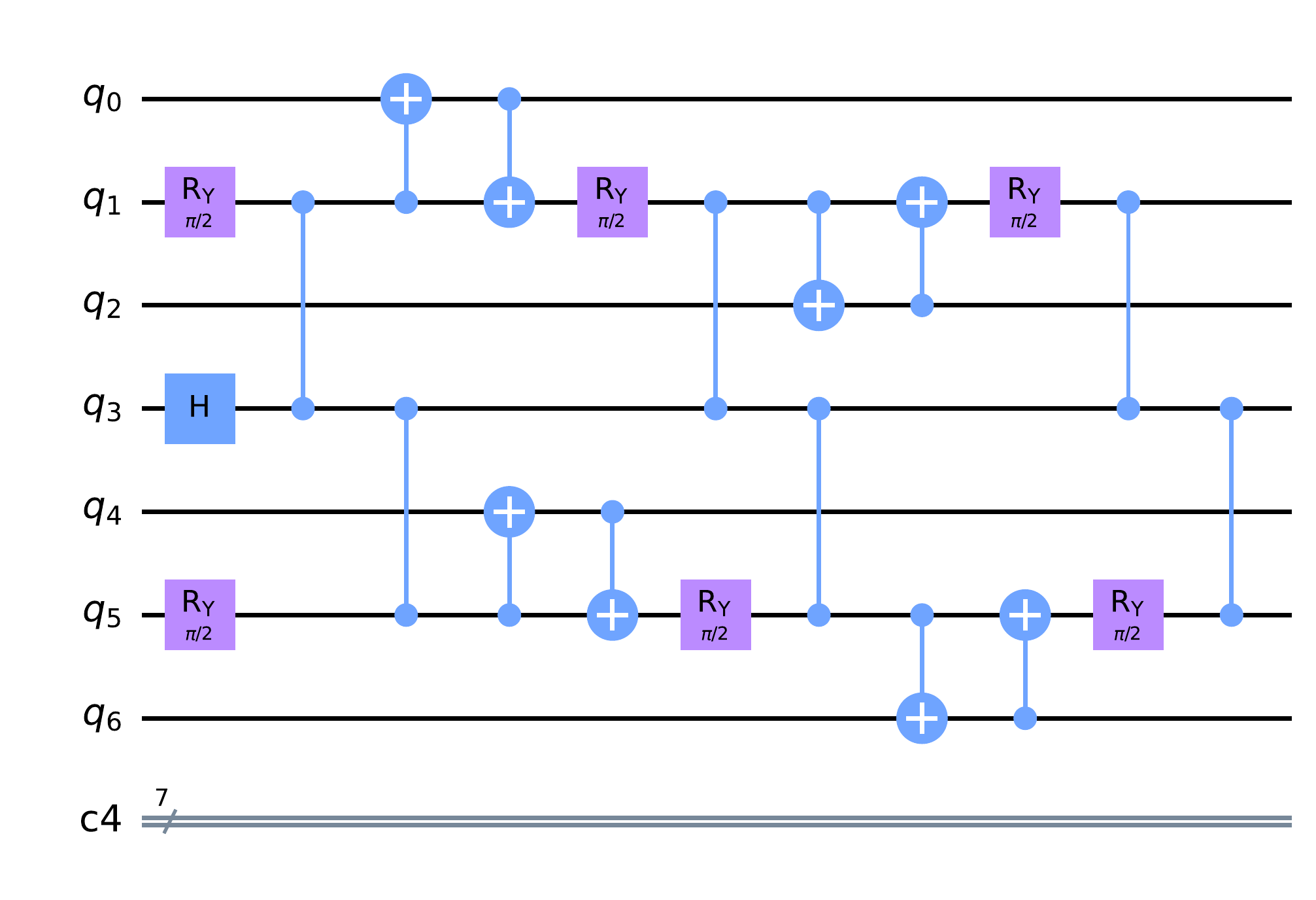}}
    \caption{
    (a) Visual representation of the process in the condensed scenario: $\mathcal{S}$ represents the system, $\mathcal{AE}$ the condensed ancilla-emitter pairs, the black lines represent the interaction between the different qubits. (b) Circuit representing the gate decomposition of the process in the condensed scenario: each line represents one of the qubits of the quantum computer. The states of the ancilla-emitter pairs are prepared with a rotation along the Y axis. The interaction between the system and each ancilla-emitter pair is a Z gate on the pair controlled by the state of the system. The circuit uses several SWAP gates (represented by two consecutive CNOTs) in order to transfer the states of the qubits in the appropriate positions. Throughout the circuit, q3 is the system qubit, all others are the ancilla-emitter pairs.}
    \label{fig:cond_cir}
\end{figure}
This operation offers a great computational advantage, considering that quantum computers are still strongly limited in the number of qubits available and, more importantly, are more susceptible to decoherence the more qubits are used in the same circuit.
What was previously a series of one and two qubit gates necessary to prepare the state of an ancilla-emitter pair is now just a single qubit gate in the condensed scenario. More specifically, the state of each ancilla-emitter pair is prepared using a $R_{y}(2\alpha)$ gate, where the angle $\alpha$ has the same meaning as in the previous case. The interaction between system and environment is once again a Z gate applied on the condensed pair controlled by the state of the system. In this case the number of SWAP gates is minimised by choosing qubit $q3$ as the system. The circuit corresponding to the model is shown in Fig.~\ref{fig:cond_cir}.
\begin{figure}
    \centering
    \includegraphics[width=0.5\textwidth]{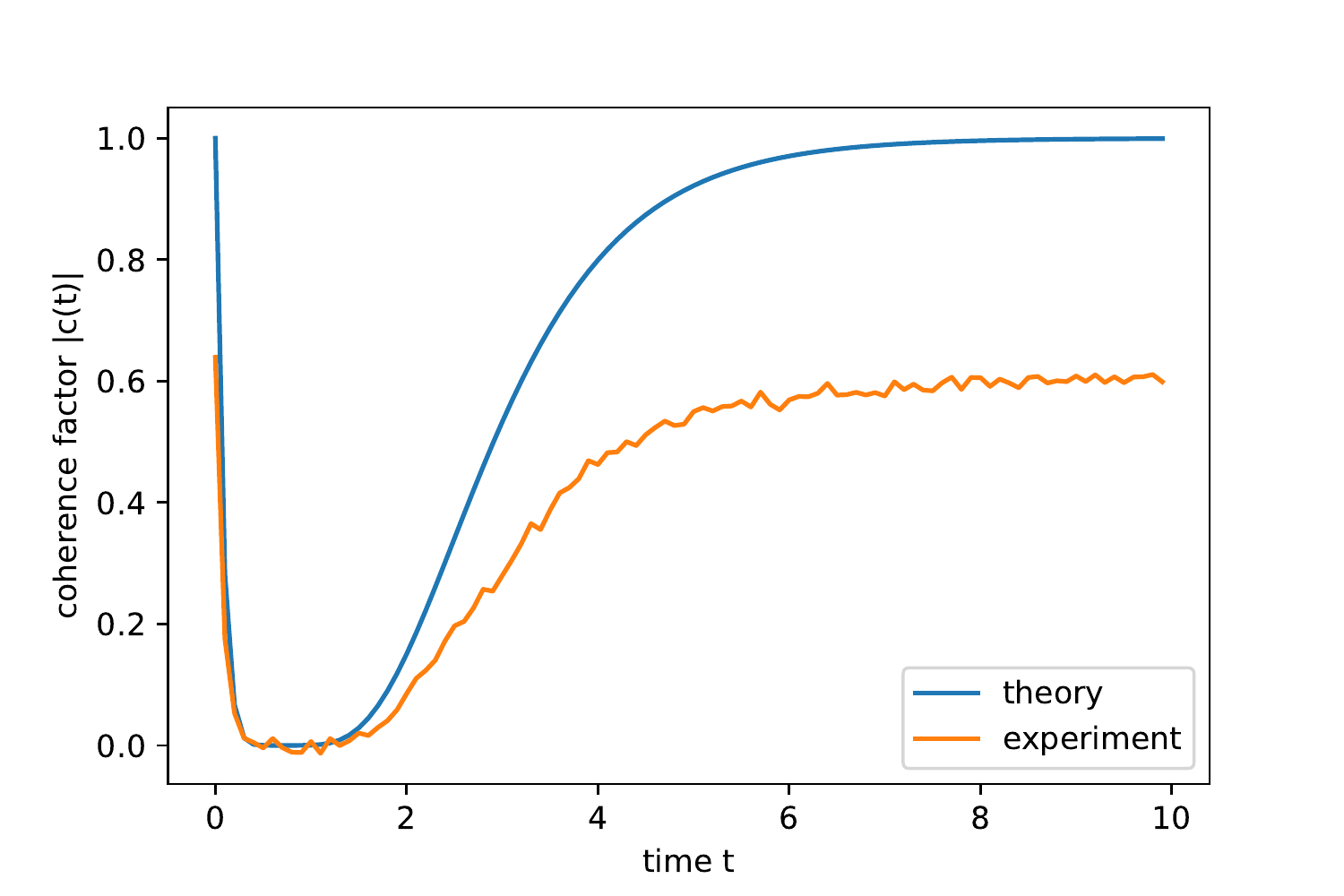}
    \caption{\textbf{Coherence factor of the system as a function of time.} Results obtained in the case of six ancilla-emitter pairs, with each pair represented by a single qubit. The coherence factor is obtained by performing state tomography at each time step. In this model the non-monotonic behaviour of the coherence factor implies a non-monotonic behaviour of the trace distance between two initially distinguishable states, and can therefore be seen as a signature of non-Markovianity.}
    \label{fig:purity}
\end{figure} 
This way, using the same quantum computer we are able to simulate six environmental emitter-ancilla pairs.

\begin{figure}
    \centering
    \includegraphics[width=0.5\textwidth]{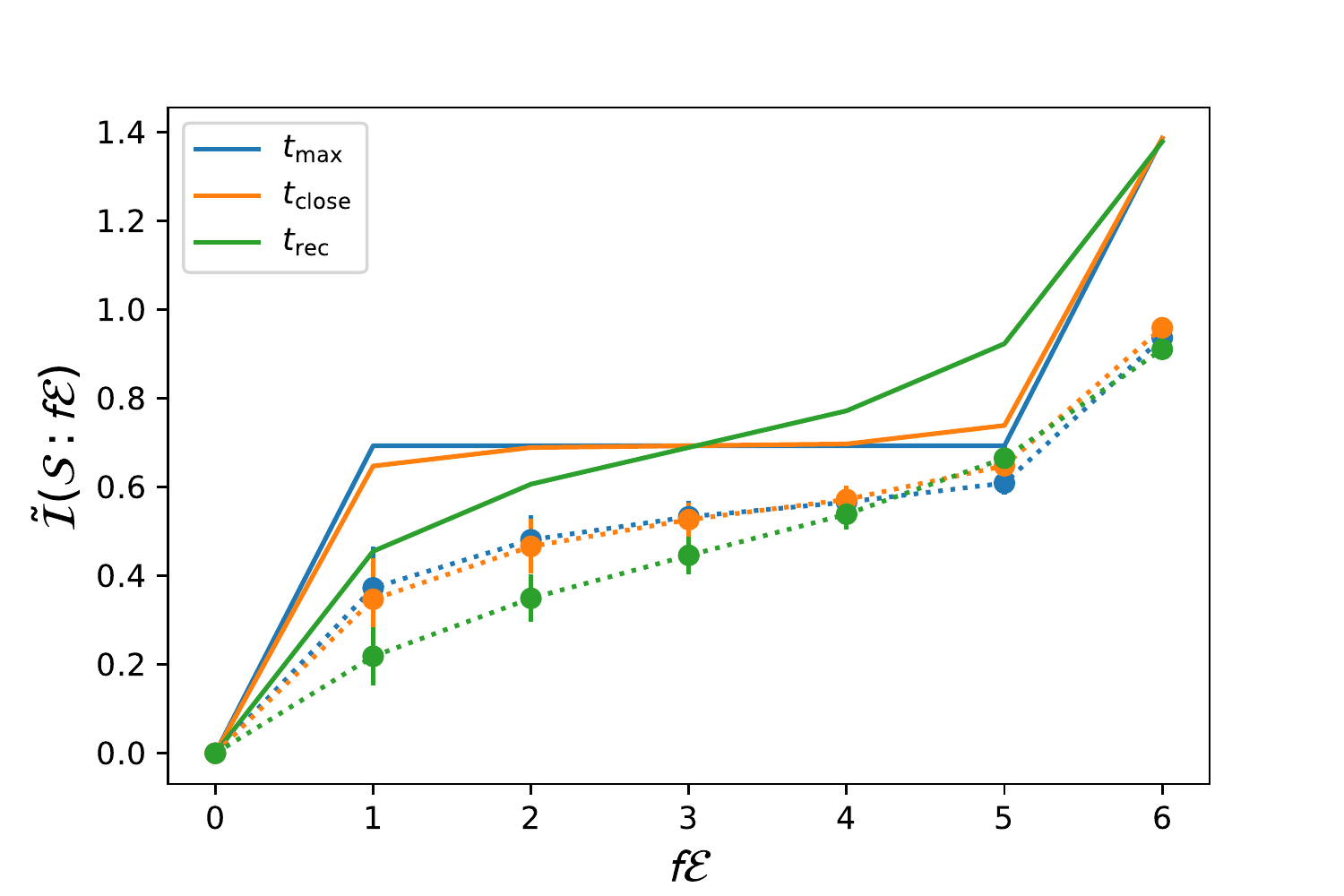}
    \caption{\textbf{Averaged mutual information between system and environmental fraction as a function of the size of the environmental fraction, for an environment of six ancilla-emitter pairs.}
    The dotted lines represent the experimental values, while the continuous lines represent the theoretical expectations. Notice how the experimental values are consistently lower than the theoretical expectations, this is due to the intrinsic decoherence the quantum computer is subject to, which erodes the mutual information between its components. 
    The three different plots correspond to three different physical times: $t=t_{\mathrm{max}}$ for which the Darwinistic features should be very sharp, a time $t=t_{\mathrm{close}}$ close to $t_{\mathrm{max}}$, and a time in the "recoherence regime" $t=t_{\mathrm{rec}}$, for which the global state should have lost its Darwinistic features. The exact values are $t_{\mathrm{max}}=\ln{2}$, $t_{\mathrm{close}}=\ln{\frac{2}{1.3}}$ and $t_{\mathrm{rec}}=\ln{6}$.}
    \label{fig:darwin}
\end{figure}

We once again run the circuit for different values of the parameter $\alpha$ corresponding to different values of the physical time and perform state tomography of the system qubit for each run of the circuit. Figure~\ref{fig:purity} shows the coherence factor of the system as a function of time, and also in this case there is a clear non-Markovian behaviour.

We recover the global state though tomographic reconstruction for three different values of the physical time. By evaluating the quantum mutual information between the system and the environmental fractions we are able to witness QD.
In Fig.~\ref{fig:darwin} we show the averaged mutual information $\Tilde{\mathcal{I}}(\mathcal{S}:f\mathcal{E})$ as a function of the environmental size $f$ for different values of the physical times. The chosen times are: $t=t_{\mathrm{max}}$ for which the Darwinistic features should be very sharp, a time $t=t_{\mathrm{close}}$ close to $t_{\mathrm{max}}$, and a time in the "recoherence regime" $t=t_{\mathrm{rec}}$, for which a significant amount of information has already flown back to the system and the global state should have lost its Darwinistic features. The exact values are $t_{\mathrm{max}}=\ln{2}$, $t_{\mathrm{close}}=\ln{\frac{2}{1.3}}$ and $t_{\mathrm{rec}}=\ln{6}$.

We can clearly see the presence of a "classical" plateau, even if lower than the theoretical expectations. Nonetheless, one of the key features of quantum Darwinism (redundancy) is reproduced, which allows us to say that, while the emergence of QD is less manifest compared to the theoretical expectations, there is a state structure clearly reminiscent of QD.

\begin{figure*}
    \centering
    \includegraphics[trim={0 2cm 0 2cm}, clip, width=\textwidth]{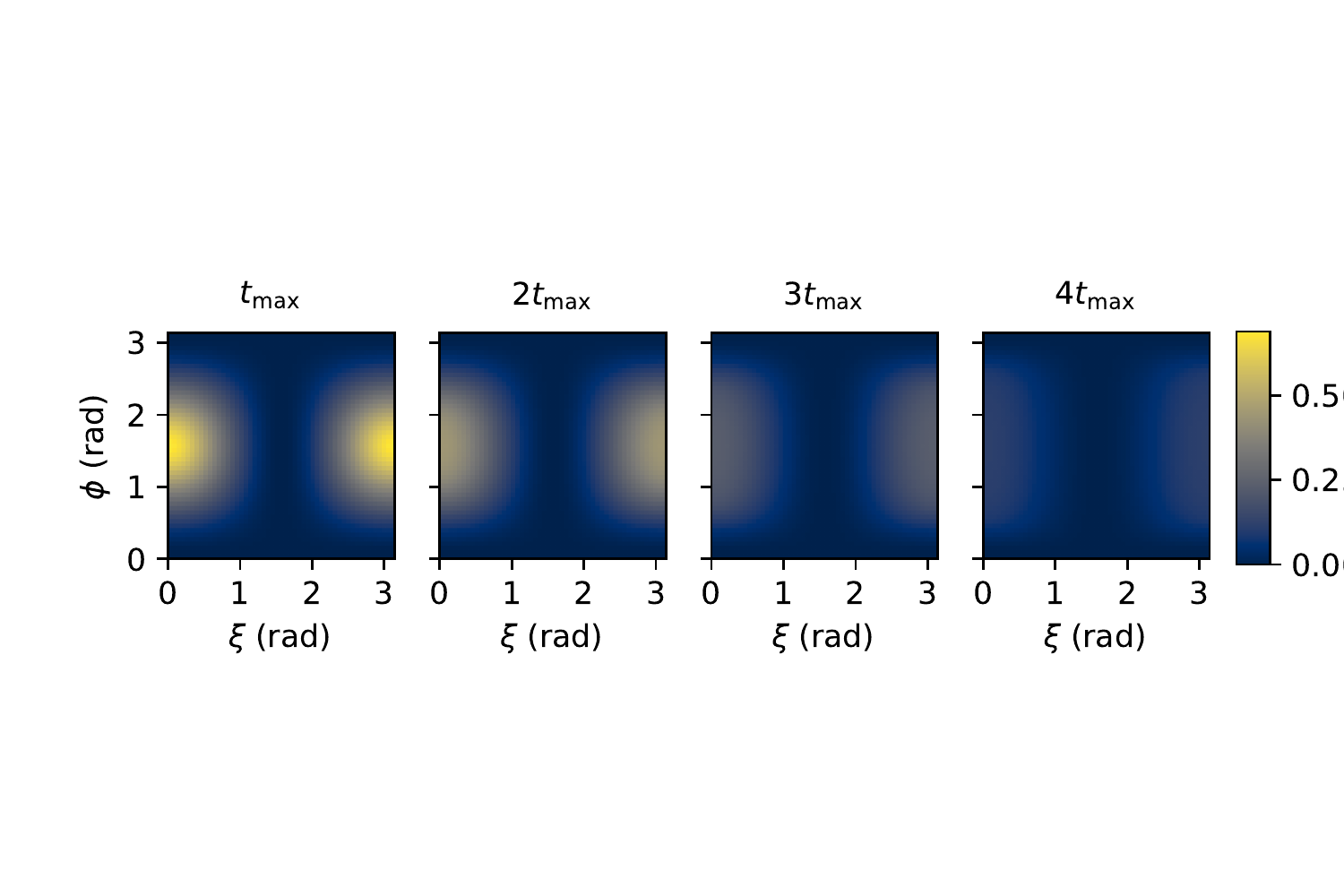}
    \caption{\textbf{Classical mutual information between the system and an ancilla-emitter pair recovered for all possible local measurement bases.} Results are obtained from the analytical solution of the model. $\phi$ and $\xi$ define a measurement basis according to $\ket{0'}=\cos(\phi/2)\ket{0}+e^{i\xi}\sin(\phi/2)\ket{1}$ and $\ket{1'}=\sin(\phi/2)\ket{0}-e^{i\xi}\cos(\phi/2)\ket{1}$. The peaks correspond to measuring the ancilla-emitter pair in the $\{\ket{+},\ \ket{-}\}$ base. The physical times are, from left to right, $t=t_{\mathrm{max}}$, $t=2t_{\mathrm{max}}$, $t=3t_{\mathrm{max}}$, $t=4t_{\mathrm{max}}$, with $t_{\mathrm{max}}=\ln 2$. The further we are from $t_{\mathrm{max}}$ the less is the obtainable CMI. }
    \label{fig:optimal}
\end{figure*}

\section{Accessible classical information}
\begin{figure*}[htb]
    \centering
    \includegraphics[trim={0 2cm 0 2cm}, clip, width=\textwidth]{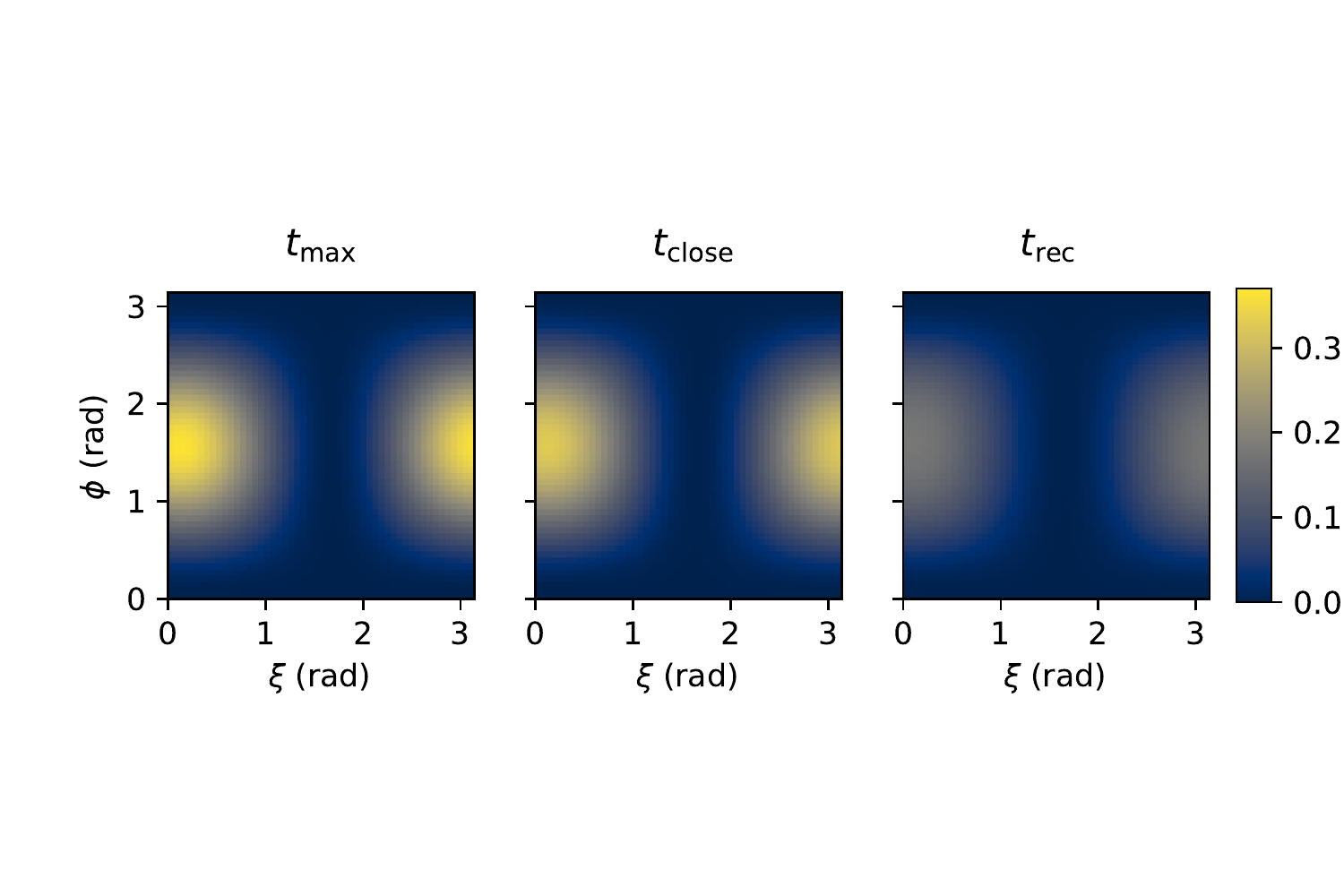}
    \caption{\textbf{Classical mutual information between the system and an ancilla-emitter pair recovered for all possible local measurement bases.} Results are obtained from the tomographically reconstructed states. $\phi$ and $\xi$ define a measurement base according to $\ket{0'}=\cos(\phi/2)\ket{0}+e^{i\xi}\sin(\phi/2)\ket{1}$ and $\ket{1'}=\sin(\phi/2)\ket{0}-e^{i\xi}\cos(\phi/2)\ket{1}$. The peaks correspond to measuring the ancilla-emitter pair in the $\{\ket{+},\ \ket{-}\}$ base. The leftmost figure corresponds to $t=t_{\mathrm{max}}$, the central to $t=t_{\mathrm{close}}$ and the one on the right is for $t=t_{\mathrm{rec}}$. Values are $t_{\mathrm{max}}=\ln{2}$, $t_{\mathrm{close}}=\ln{\frac{2}{1.3}}$ and $t_{\mathrm{rec}}=\ln{6}$.}
    \label{fig:real_optimal}
\end{figure*}
Quantum Darwinism tells us the necessary conditions to achieve system objectivity. The question arises however over how much information can actually be inferred on the system by observers making measurements of the environment.

The statistical correlations between measurements performed on the environmental fractions and measurements performed on the system in its pointer basis are bounded by the Holevo bound computed in the pointer basis of the system~\cite{holevo1973bounds}. The Holevo bound maximised over the measurement basis of the system is numerically the same as the QMI only if there is zero discord between the system and the environmental fractions. Representing information non-locally accessible, it has been suggested~\cite{PhysRevLett.112.120402, PhysRevA.91.032122, PhysRevLett.122.010403, touil2021eavesdropping, e23080995} that discord should be zero in objective states, leading to the formulation of alternative ways to witness objectivity (albeit sharing the same core ideas of QD): Spectrum Broadcast Structure~\cite{PhysRevLett.112.120402, PhysRevA.91.032122} and strong quantum Darwinism~\cite{PhysRevLett.122.010403}.

The Holevo bound is however only a bound. It does not hold information over what actions an observer would actually have to perform in order to obtain information about the system.

As we are interested in assessing the information accessible through possible measurements on the environment, we choose to evaluate the classical mutual information (CMI) between the system and the environment. Measurements on the system are always performed in the pointer basis, which happens to be the computational one for this particular model. We aim at finding the environmental measurement basis maximising the CMI between the system and one ancilla-emitter pair. The measurement basis maximising CMI between the system and two or more ancilla-emitter pairs may as well be different, but here we want to maximise what an observer can infer from just one pair. We will also restrict ourselves to the case of projective measurements.

In Fig.~\ref{fig:optimal} we show the CMI between system and one ancilla-emitter pair for all possible one qubit measurement bases, calculated using the analytical solution of the model. On the horizontal and vertical axis there are two angles $\phi$ and $\xi$ that uniquely define a single qubit basis defined as $\ket{0'}=\cos(\phi/2)\ket{0}+e^{i\xi}\sin(\phi/2)\ket{1}$ and $\ket{1'}=\sin(\phi/2)\ket{0}-e^{i\xi}\cos(\phi/2)\ket{1}$.

\begin{figure}
    \centering
    \includegraphics[width=\textwidth]{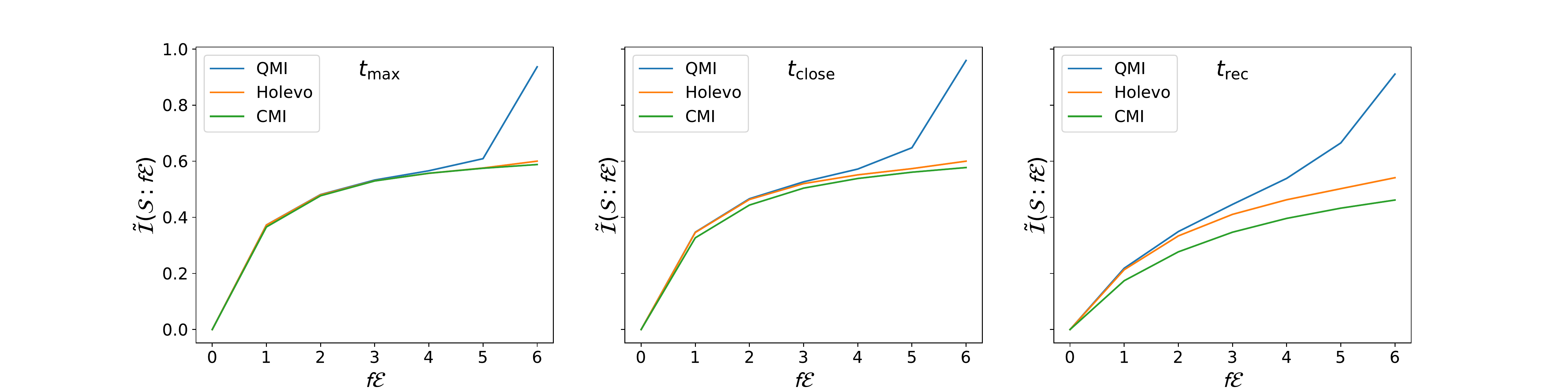}
    \caption{\textbf{Comparison between quantum mutual information, classical mutual information and Holevo bound for the three tomographically reconstructed states.} While the three quantities have the same values for $t_{\mathrm{max}}$, there are discrepancies for the other physical times, in particular, CMI is consistently lower than the Holevo bound, which is in turn consistently lower than the QMI. The leftmost figure corresponds to $t=t_{\mathrm{max}}$, the central to $t=t_{\mathrm{close}}$ and the one on the right is for $t=t_{\mathrm{rec}}$. Values are $t_{\mathrm{max}}=\ln{2}$, $t_{\mathrm{close}}=\ln{\frac{2}{1.3}}$ and $t_{\mathrm{rec}}=\ln{6}$.}
    \label{fig:classic and darwin}
\end{figure}
In Fig.~\ref{fig:real_optimal} we show the same quantity by using the experimental states reconstructed using tomography.
Both figures show that the classical mutual information is maximised by measuring the environmental qubits in the $\{\ket{+},\ \ket{-}\}$ basis.\\
\indent Finally, by imposing that the environment is measured in the same measurement basis for all the environmental qubits, we compute the CMI between the system and an environmental fraction as a function of the environmental size.
We show in Fig.~\ref{fig:classic and darwin} a comparison between the QMI, the Holevo bound and the maximised CMI, we can see that, when $t=t_{\mathrm{max}}$ and $t=t_{\mathrm{close}}$, the CMI can be used as a signature of objectivity.
It is also interesting to notice that while the Holevo bound and the maximised CMI are almost the same at $t=t_{\mathrm{max}}$ and $t=t_{\mathrm{close}}$, they are instead numerically different at $t=t_{\mathrm{rec}}$, meaning that when the state is far from being objective, it is not possible to reach the Holevo bound through local projective measurements alone.

\section{Conclusions}

In this work we have witnessed objectivity in the framework of quantum Darwinism, for a stochastic collision model, using a quantum computer as the experimental platform; which paves the way for future applications of quantum computers to foundation problems such as the definition of objectivity and the quantum to classical transition.

Our results show that, with the appropriate simulation techniques, current NISQ devices are capable of reproducing states with objective features. The quantum computer is subject to uncontrolled decoherence due to its own surrounding environment (with an irreversible loss of quantum information) that, along with the crosstalk between qubits hinders the emergence of objectivity. Nonetheless it is reasonable to assume, given the rapid technological advance of the field, that it will soon be possible to produce with high fidelity states that cannot be currently modeled with classical resources alone. However, it is important to point out that the results presented here require to perform the full tomographic reconstruction of the global state. Regardless of the level of technological advancements, full tomography is an expensive task that scales exponentially with the size of the system, and represent the biggest bottleneck in terms of experimental realization of large objective states. This motivates the search for appropriate and efficient witnesses of objectivity, an interesting future direction in the field of quantum Darwinism.
\section{Acknowledgements}

We acknowledge the use of IBM Quantum services for this work. The views expressed are those of the authors, and do not reflect the official policy or position of IBM or the IBM Quantum team. D.A.C. and G.M.P. acknowledge support from MIUR through project PRIN (Project No. 2017SRN-BRK QUSHIP). G.G.-P., M.A.C.R., and S.M. acknowledge financial support from the Academy of Finland via the Centre of Excellence program (Project no. 312058). S.M. and G.G.-P. acknowledge support from the emmy.network foundation under the aegis of the Fondation de Luxembourg. G.G.-P. acknowledges support from the Academy of Finland via the Postdoctoral Researcher program (Project no. 341985). The computer resources of the Finnish IT Center for Science (CSC) and the FGCI project (Finland) are acknowledged.

\bibliography{Reference}

\begin{thebibliography}{43}%
\makeatletter
\providecommand \@ifxundefined [1]{%
 \@ifx{#1\undefined}
}%
\providecommand \@ifnum [1]{%
 \ifnum #1\expandafter \@firstoftwo
 \else \expandafter \@secondoftwo
 \fi
}%
\providecommand \@ifx [1]{%
 \ifx #1\expandafter \@firstoftwo
 \else \expandafter \@secondoftwo
 \fi
}%
\providecommand \natexlab [1]{#1}%
\providecommand \enquote  [1]{``#1''}%
\providecommand \bibnamefont  [1]{#1}%
\providecommand \bibfnamefont [1]{#1}%
\providecommand \citenamefont [1]{#1}%
\providecommand \href@noop [0]{\@secondoftwo}%
\providecommand \href [0]{\begingroup \@sanitize@url \@href}%
\providecommand \@href[1]{\@@startlink{#1}\@@href}%
\providecommand \@@href[1]{\endgroup#1\@@endlink}%
\providecommand \@sanitize@url [0]{\catcode `\\12\catcode `\$12\catcode
  `\&12\catcode `\#12\catcode `\^12\catcode `\_12\catcode `\%12\relax}%
\providecommand \@@startlink[1]{}%
\providecommand \@@endlink[0]{}%
\providecommand \url  [0]{\begingroup\@sanitize@url \@url }%
\providecommand \@url [1]{\endgroup\@href {#1}{\urlprefix }}%
\providecommand \urlprefix  [0]{URL }%
\providecommand \Eprint [0]{\href }%
\providecommand \doibase [0]{http://dx.doi.org/}%
\providecommand \selectlanguage [0]{\@gobble}%
\providecommand \bibinfo  [0]{\@secondoftwo}%
\providecommand \bibfield  [0]{\@secondoftwo}%
\providecommand \translation [1]{[#1]}%
\providecommand \BibitemOpen [0]{}%
\providecommand \bibitemStop [0]{}%
\providecommand \bibitemNoStop [0]{.\EOS\space}%
\providecommand \EOS [0]{\spacefactor3000\relax}%
\providecommand \BibitemShut  [1]{\csname bibitem#1\endcsname}%
\let\auto@bib@innerbib\@empty
\bibitem [{\citenamefont {Zurek}(2003)}]{RevModPhys.75.715}%
  \BibitemOpen
  \bibfield  {author} {\bibinfo {author} {\bibfnamefont {Wojciech~Hubert}\
  \bibnamefont {Zurek}},\ }\bibfield  {title} {\enquote {\bibinfo {title}
  {Decoherence, einselection, and the quantum origins of the classical},}\
  }\href {\doibase 10.1103/RevModPhys.75.715} {\bibfield  {journal} {\bibinfo
  {journal} {Rev. Mod. Phys.}\ }\textbf {\bibinfo {volume} {75}},\ \bibinfo
  {pages} {715--775} (\bibinfo {year} {2003})}\BibitemShut {NoStop}%
\bibitem [{\citenamefont {Zurek}(1981)}]{PhysRevD.24.1516}%
  \BibitemOpen
  \bibfield  {author} {\bibinfo {author} {\bibfnamefont {W.~H.}\ \bibnamefont
  {Zurek}},\ }\bibfield  {title} {\enquote {\bibinfo {title} {Pointer basis of
  quantum apparatus: Into what mixture does the wave packet collapse?}}\ }\href
  {\doibase 10.1103/PhysRevD.24.1516} {\bibfield  {journal} {\bibinfo
  {journal} {Phys. Rev. D}\ }\textbf {\bibinfo {volume} {24}},\ \bibinfo
  {pages} {1516--1525} (\bibinfo {year} {1981})}\BibitemShut {NoStop}%
\bibitem [{\citenamefont {Zurek}(2009)}]{Zurek2009}%
  \BibitemOpen
  \bibfield  {author} {\bibinfo {author} {\bibfnamefont {Wojciech~Hubert}\
  \bibnamefont {Zurek}},\ }\bibfield  {title} {\enquote {\bibinfo {title}
  {Quantum darwinism},}\ }\href {\doibase 10.1038/nphys1202} {\bibfield
  {journal} {\bibinfo  {journal} {Nature Physics}\ }\textbf {\bibinfo {volume}
  {5}},\ \bibinfo {pages} {181--188} (\bibinfo {year} {2009})}\BibitemShut
  {NoStop}%
\bibitem [{\citenamefont {Blume-Kohout}\ and\ \citenamefont
  {Zurek}(2006)}]{PhysRevA.73.062310}%
  \BibitemOpen
  \bibfield  {author} {\bibinfo {author} {\bibfnamefont {Robin}\ \bibnamefont
  {Blume-Kohout}}\ and\ \bibinfo {author} {\bibfnamefont {Wojciech~H.}\
  \bibnamefont {Zurek}},\ }\bibfield  {title} {\enquote {\bibinfo {title}
  {Quantum darwinism: Entanglement, branches, and the emergent classicality of
  redundantly stored quantum information},}\ }\href {\doibase
  10.1103/PhysRevA.73.062310} {\bibfield  {journal} {\bibinfo  {journal} {Phys.
  Rev. A}\ }\textbf {\bibinfo {volume} {73}},\ \bibinfo {pages} {062310}
  (\bibinfo {year} {2006})}\BibitemShut {NoStop}%
\bibitem [{\citenamefont {Breuer}\ \emph {et~al.}(2002)\citenamefont {Breuer},
  \citenamefont {Petruccione} \emph {et~al.}}]{breuer2002theory}%
  \BibitemOpen
  \bibfield  {author} {\bibinfo {author} {\bibfnamefont {Heinz-Peter}\
  \bibnamefont {Breuer}}, \bibinfo {author} {\bibfnamefont {Francesco}\
  \bibnamefont {Petruccione}},  \emph {et~al.},\ }\href@noop {} {\emph
  {\bibinfo {title} {The theory of open quantum systems}}}\ (\bibinfo
  {publisher} {Oxford University Press on Demand},\ \bibinfo {year}
  {2002})\BibitemShut {NoStop}%
\bibitem [{\citenamefont {Rivas}\ \emph {et~al.}(2014)\citenamefont {Rivas},
  \citenamefont {Huelga},\ and\ \citenamefont {Plenio}}]{Rivas_2014}%
  \BibitemOpen
  \bibfield  {author} {\bibinfo {author} {\bibfnamefont {{\'{A}}ngel}\
  \bibnamefont {Rivas}}, \bibinfo {author} {\bibfnamefont {Susana~F}\
  \bibnamefont {Huelga}}, \ and\ \bibinfo {author} {\bibfnamefont {Martin~B}\
  \bibnamefont {Plenio}},\ }\bibfield  {title} {\enquote {\bibinfo {title}
  {Quantum non-markovianity: characterization, quantification and detection},}\
  }\href {\doibase 10.1088/0034-4885/77/9/094001} {\bibfield  {journal}
  {\bibinfo  {journal} {Reports on Progress in Physics}\ }\textbf {\bibinfo
  {volume} {77}},\ \bibinfo {pages} {094001} (\bibinfo {year}
  {2014})}\BibitemShut {NoStop}%
\bibitem [{\citenamefont {Auff{\`e}ves}\ and\ \citenamefont
  {Grangier}(2016)}]{Auffeves2016}%
  \BibitemOpen
  \bibfield  {author} {\bibinfo {author} {\bibfnamefont {Alexia}\ \bibnamefont
  {Auff{\`e}ves}}\ and\ \bibinfo {author} {\bibfnamefont {Philippe}\
  \bibnamefont {Grangier}},\ }\bibfield  {title} {\enquote {\bibinfo {title}
  {Contexts, systems and modalities: A new ontology for quantum mechanics},}\
  }\href {\doibase 10.1007/s10701-015-9952-z} {\bibfield  {journal} {\bibinfo
  {journal} {Foundations of Physics}\ }\textbf {\bibinfo {volume} {46}},\
  \bibinfo {pages} {121--137} (\bibinfo {year} {2016})}\BibitemShut {NoStop}%
\bibitem [{\citenamefont {Blume-Kohout}\ and\ \citenamefont
  {Zurek}(2005)}]{Blume-Kohout2005}%
  \BibitemOpen
  \bibfield  {author} {\bibinfo {author} {\bibfnamefont {Robin}\ \bibnamefont
  {Blume-Kohout}}\ and\ \bibinfo {author} {\bibfnamefont {Wojciech~H.}\
  \bibnamefont {Zurek}},\ }\bibfield  {title} {\enquote {\bibinfo {title} {A
  simple example of ``quantum darwinism'': Redundant information storage in
  many-spin environments},}\ }\href {\doibase 10.1007/s10701-005-7352-5}
  {\bibfield  {journal} {\bibinfo  {journal} {Foundations of Physics}\ }\textbf
  {\bibinfo {volume} {35}},\ \bibinfo {pages} {1857--1876} (\bibinfo {year}
  {2005})}\BibitemShut {NoStop}%
\bibitem [{\citenamefont {Ciampini}\ \emph {et~al.}(2018)\citenamefont
  {Ciampini}, \citenamefont {Pinna}, \citenamefont {Mataloni},\ and\
  \citenamefont {Paternostro}}]{PhysRevA.98.020101}%
  \BibitemOpen
  \bibfield  {author} {\bibinfo {author} {\bibfnamefont {Mario~A.}\
  \bibnamefont {Ciampini}}, \bibinfo {author} {\bibfnamefont {Giorgia}\
  \bibnamefont {Pinna}}, \bibinfo {author} {\bibfnamefont {Paolo}\ \bibnamefont
  {Mataloni}}, \ and\ \bibinfo {author} {\bibfnamefont {Mauro}\ \bibnamefont
  {Paternostro}},\ }\bibfield  {title} {\enquote {\bibinfo {title}
  {Experimental signature of quantum darwinism in photonic cluster states},}\
  }\href {\doibase 10.1103/PhysRevA.98.020101} {\bibfield  {journal} {\bibinfo
  {journal} {Phys. Rev. A}\ }\textbf {\bibinfo {volume} {98}},\ \bibinfo
  {pages} {020101} (\bibinfo {year} {2018})}\BibitemShut {NoStop}%
\bibitem [{\citenamefont {Unden}\ \emph {et~al.}(2019)\citenamefont {Unden},
  \citenamefont {Louzon}, \citenamefont {Zwolak}, \citenamefont {Zurek},\ and\
  \citenamefont {Jelezko}}]{PhysRevLett.123.140402}%
  \BibitemOpen
  \bibfield  {author} {\bibinfo {author} {\bibfnamefont {T.~K.}\ \bibnamefont
  {Unden}}, \bibinfo {author} {\bibfnamefont {D.}~\bibnamefont {Louzon}},
  \bibinfo {author} {\bibfnamefont {M.}~\bibnamefont {Zwolak}}, \bibinfo
  {author} {\bibfnamefont {W.~H.}\ \bibnamefont {Zurek}}, \ and\ \bibinfo
  {author} {\bibfnamefont {F.}~\bibnamefont {Jelezko}},\ }\bibfield  {title}
  {\enquote {\bibinfo {title} {Revealing the emergence of classicality using
  nitrogen-vacancy centers},}\ }\href {\doibase 10.1103/PhysRevLett.123.140402}
  {\bibfield  {journal} {\bibinfo  {journal} {Phys. Rev. Lett.}\ }\textbf
  {\bibinfo {volume} {123}},\ \bibinfo {pages} {140402} (\bibinfo {year}
  {2019})}\BibitemShut {NoStop}%
\bibitem [{\citenamefont {Chen}\ \emph {et~al.}(2019)\citenamefont {Chen},
  \citenamefont {Zhong}, \citenamefont {Li}, \citenamefont {Wu}, \citenamefont
  {Wang}, \citenamefont {Li}, \citenamefont {Liu}, \citenamefont {Lu},\ and\
  \citenamefont {Pan}}]{CHEN2019580}%
  \BibitemOpen
  \bibfield  {author} {\bibinfo {author} {\bibfnamefont {Ming-Cheng}\
  \bibnamefont {Chen}}, \bibinfo {author} {\bibfnamefont {Han-Sen}\
  \bibnamefont {Zhong}}, \bibinfo {author} {\bibfnamefont {Yuan}\ \bibnamefont
  {Li}}, \bibinfo {author} {\bibfnamefont {Dian}\ \bibnamefont {Wu}}, \bibinfo
  {author} {\bibfnamefont {Xi-Lin}\ \bibnamefont {Wang}}, \bibinfo {author}
  {\bibfnamefont {Li}~\bibnamefont {Li}}, \bibinfo {author} {\bibfnamefont
  {Nai-Le}\ \bibnamefont {Liu}}, \bibinfo {author} {\bibfnamefont {Chao-Yang}\
  \bibnamefont {Lu}}, \ and\ \bibinfo {author} {\bibfnamefont {Jian-Wei}\
  \bibnamefont {Pan}},\ }\bibfield  {title} {\enquote {\bibinfo {title}
  {Emergence of classical objectivity of quantum darwinism in a photonic
  quantum simulator},}\ }\href {\doibase
  https://doi.org/10.1016/j.scib.2019.03.032} {\bibfield  {journal} {\bibinfo
  {journal} {Science Bulletin}\ }\textbf {\bibinfo {volume} {64}},\ \bibinfo
  {pages} {580--585} (\bibinfo {year} {2019})}\BibitemShut {NoStop}%
\bibitem [{\citenamefont {Riedel}\ \emph {et~al.}(2012)\citenamefont {Riedel},
  \citenamefont {Zurek},\ and\ \citenamefont {Zwolak}}]{Jess_Riedel_2012}%
  \BibitemOpen
  \bibfield  {author} {\bibinfo {author} {\bibfnamefont {C~Jess}\ \bibnamefont
  {Riedel}}, \bibinfo {author} {\bibfnamefont {Wojciech~H}\ \bibnamefont
  {Zurek}}, \ and\ \bibinfo {author} {\bibfnamefont {Michael}\ \bibnamefont
  {Zwolak}},\ }\bibfield  {title} {\enquote {\bibinfo {title} {The rise and
  fall of redundancy in decoherence and quantum darwinism},}\ }\href {\doibase
  10.1088/1367-2630/14/8/083010} {\bibfield  {journal} {\bibinfo  {journal}
  {New Journal of Physics}\ }\textbf {\bibinfo {volume} {14}},\ \bibinfo
  {pages} {083010} (\bibinfo {year} {2012})}\BibitemShut {NoStop}%
\bibitem [{\citenamefont {Giorgi}\ \emph {et~al.}(2015)\citenamefont {Giorgi},
  \citenamefont {Galve},\ and\ \citenamefont {Zambrini}}]{PhysRevA.92.022105}%
  \BibitemOpen
  \bibfield  {author} {\bibinfo {author} {\bibfnamefont {Gian~Luca}\
  \bibnamefont {Giorgi}}, \bibinfo {author} {\bibfnamefont {Fernando}\
  \bibnamefont {Galve}}, \ and\ \bibinfo {author} {\bibfnamefont {Roberta}\
  \bibnamefont {Zambrini}},\ }\bibfield  {title} {\enquote {\bibinfo {title}
  {Quantum darwinism and non-markovian dissipative dynamics from quantum phases
  of the spin-1/2 $xx$ model},}\ }\href {\doibase 10.1103/PhysRevA.92.022105}
  {\bibfield  {journal} {\bibinfo  {journal} {Phys. Rev. A}\ }\textbf {\bibinfo
  {volume} {92}},\ \bibinfo {pages} {022105} (\bibinfo {year}
  {2015})}\BibitemShut {NoStop}%
\bibitem [{\citenamefont {Lampo}\ \emph {et~al.}(2017)\citenamefont {Lampo},
  \citenamefont {Tuziemski}, \citenamefont {Lewenstein},\ and\ \citenamefont
  {Korbicz}}]{PhysRevA.96.012120}%
  \BibitemOpen
  \bibfield  {author} {\bibinfo {author} {\bibfnamefont {Aniello}\ \bibnamefont
  {Lampo}}, \bibinfo {author} {\bibfnamefont {Jan}\ \bibnamefont {Tuziemski}},
  \bibinfo {author} {\bibfnamefont {Maciej}\ \bibnamefont {Lewenstein}}, \ and\
  \bibinfo {author} {\bibfnamefont {Jaros\l{}aw~K.}\ \bibnamefont {Korbicz}},\
  }\bibfield  {title} {\enquote {\bibinfo {title} {Objectivity in the
  non-markovian spin-boson model},}\ }\href {\doibase
  10.1103/PhysRevA.96.012120} {\bibfield  {journal} {\bibinfo  {journal} {Phys.
  Rev. A}\ }\textbf {\bibinfo {volume} {96}},\ \bibinfo {pages} {012120}
  (\bibinfo {year} {2017})}\BibitemShut {NoStop}%
\bibitem [{\citenamefont {Pleasance}\ and\ \citenamefont
  {Garraway}(2017)}]{PhysRevA.96.062105}%
  \BibitemOpen
  \bibfield  {author} {\bibinfo {author} {\bibfnamefont {Graeme}\ \bibnamefont
  {Pleasance}}\ and\ \bibinfo {author} {\bibfnamefont {Barry~M.}\ \bibnamefont
  {Garraway}},\ }\bibfield  {title} {\enquote {\bibinfo {title} {Application of
  quantum darwinism to a structured environment},}\ }\href {\doibase
  10.1103/PhysRevA.96.062105} {\bibfield  {journal} {\bibinfo  {journal} {Phys.
  Rev. A}\ }\textbf {\bibinfo {volume} {96}},\ \bibinfo {pages} {062105}
  (\bibinfo {year} {2017})}\BibitemShut {NoStop}%
\bibitem [{\citenamefont {Campbell}\ \emph {et~al.}(2019)\citenamefont
  {Campbell}, \citenamefont {\ifmmode~\mbox{\c{C}}\else \c{C}\fi{}akmak},
  \citenamefont {M\"ustecapl\ifmmode \imath \else \i
  \fi{}o\ifmmode~\breve{g}\else \u{g}\fi{}lu}, \citenamefont {Paternostro},\
  and\ \citenamefont {Vacchini}}]{PhysRevA.99.042103}%
  \BibitemOpen
  \bibfield  {author} {\bibinfo {author} {\bibfnamefont {Steve}\ \bibnamefont
  {Campbell}}, \bibinfo {author} {\bibfnamefont {Bar\ifmmode \imath \else \i
  \fi{}\ifmmode \mbox{\c{s}}\else~\c{s}\fi{}}\ \bibnamefont
  {\ifmmode~\mbox{\c{C}}\else \c{C}\fi{}akmak}}, \bibinfo {author}
  {\bibfnamefont {\"Ozg\"ur~E.}\ \bibnamefont {M\"ustecapl\ifmmode \imath \else
  \i \fi{}o\ifmmode~\breve{g}\else \u{g}\fi{}lu}}, \bibinfo {author}
  {\bibfnamefont {Mauro}\ \bibnamefont {Paternostro}}, \ and\ \bibinfo {author}
  {\bibfnamefont {Bassano}\ \bibnamefont {Vacchini}},\ }\bibfield  {title}
  {\enquote {\bibinfo {title} {Collisional unfolding of quantum darwinism},}\
  }\href {\doibase 10.1103/PhysRevA.99.042103} {\bibfield  {journal} {\bibinfo
  {journal} {Phys. Rev. A}\ }\textbf {\bibinfo {volume} {99}},\ \bibinfo
  {pages} {042103} (\bibinfo {year} {2019})}\BibitemShut {NoStop}%
\bibitem [{\citenamefont {Milazzo}\ \emph {et~al.}(2019)\citenamefont
  {Milazzo}, \citenamefont {Lorenzo}, \citenamefont {Paternostro},\ and\
  \citenamefont {Palma}}]{PhysRevA.100.012101}%
  \BibitemOpen
  \bibfield  {author} {\bibinfo {author} {\bibfnamefont {Nadia}\ \bibnamefont
  {Milazzo}}, \bibinfo {author} {\bibfnamefont {Salvatore}\ \bibnamefont
  {Lorenzo}}, \bibinfo {author} {\bibfnamefont {Mauro}\ \bibnamefont
  {Paternostro}}, \ and\ \bibinfo {author} {\bibfnamefont {G.~Massimo}\
  \bibnamefont {Palma}},\ }\bibfield  {title} {\enquote {\bibinfo {title} {Role
  of information backflow in the emergence of quantum darwinism},}\ }\href
  {\doibase 10.1103/PhysRevA.100.012101} {\bibfield  {journal} {\bibinfo
  {journal} {Phys. Rev. A}\ }\textbf {\bibinfo {volume} {100}},\ \bibinfo
  {pages} {012101} (\bibinfo {year} {2019})}\BibitemShut {NoStop}%
\bibitem [{\citenamefont {Oliveira}\ \emph {et~al.}(2019)\citenamefont
  {Oliveira}, \citenamefont {de~Paula},\ and\ \citenamefont
  {Drumond}}]{PhysRevA.100.052110}%
  \BibitemOpen
  \bibfield  {author} {\bibinfo {author} {\bibfnamefont {S.~M.}\ \bibnamefont
  {Oliveira}}, \bibinfo {author} {\bibfnamefont {A.~L.}\ \bibnamefont
  {de~Paula}}, \ and\ \bibinfo {author} {\bibfnamefont {R.~C.}\ \bibnamefont
  {Drumond}},\ }\bibfield  {title} {\enquote {\bibinfo {title} {Quantum
  darwinism and non-markovianity in a model of quantum harmonic oscillators},}\
  }\href {\doibase 10.1103/PhysRevA.100.052110} {\bibfield  {journal} {\bibinfo
   {journal} {Phys. Rev. A}\ }\textbf {\bibinfo {volume} {100}},\ \bibinfo
  {pages} {052110} (\bibinfo {year} {2019})}\BibitemShut {NoStop}%
\bibitem [{\citenamefont {Lorenzo}\ \emph {et~al.}(2020)\citenamefont
  {Lorenzo}, \citenamefont {Paternostro},\ and\ \citenamefont
  {Palma}}]{PhysRevResearch.2.013164}%
  \BibitemOpen
  \bibfield  {author} {\bibinfo {author} {\bibfnamefont {Salvatore}\
  \bibnamefont {Lorenzo}}, \bibinfo {author} {\bibfnamefont {Mauro}\
  \bibnamefont {Paternostro}}, \ and\ \bibinfo {author} {\bibfnamefont
  {G.~Massimo}\ \bibnamefont {Palma}},\ }\bibfield  {title} {\enquote {\bibinfo
  {title} {Anti-zeno-based dynamical control of the unfolding of quantum
  darwinism},}\ }\href {\doibase 10.1103/PhysRevResearch.2.013164} {\bibfield
  {journal} {\bibinfo  {journal} {Phys. Rev. Research}\ }\textbf {\bibinfo
  {volume} {2}},\ \bibinfo {pages} {013164} (\bibinfo {year}
  {2020})}\BibitemShut {NoStop}%
\bibitem [{\citenamefont {Le}\ and\ \citenamefont
  {Olaya-Castro}(2020)}]{Le_2020}%
  \BibitemOpen
  \bibfield  {author} {\bibinfo {author} {\bibfnamefont {Thao~P}\ \bibnamefont
  {Le}}\ and\ \bibinfo {author} {\bibfnamefont {Alexandra}\ \bibnamefont
  {Olaya-Castro}},\ }\bibfield  {title} {\enquote {\bibinfo {title} {Witnessing
  non-objectivity in the framework of strong quantum darwinism},}\ }\href
  {\doibase 10.1088/2058-9565/abac4e} {\bibfield  {journal} {\bibinfo
  {journal} {Quantum Science and Technology}\ }\textbf {\bibinfo {volume}
  {5}},\ \bibinfo {pages} {045012} (\bibinfo {year} {2020})}\BibitemShut
  {NoStop}%
\bibitem [{\citenamefont {Çakmak}\ \emph {et~al.}(2021)\citenamefont
  {Çakmak}, \citenamefont {Müstecaplıoğlu}, \citenamefont {Paternostro},
  \citenamefont {Vacchini},\ and\ \citenamefont {Campbell}}]{e23080995}%
  \BibitemOpen
  \bibfield  {author} {\bibinfo {author} {\bibfnamefont {Barış}\ \bibnamefont
  {Çakmak}}, \bibinfo {author} {\bibfnamefont {Özgür~E.}\ \bibnamefont
  {Müstecaplıoğlu}}, \bibinfo {author} {\bibfnamefont {Mauro}\ \bibnamefont
  {Paternostro}}, \bibinfo {author} {\bibfnamefont {Bassano}\ \bibnamefont
  {Vacchini}}, \ and\ \bibinfo {author} {\bibfnamefont {Steve}\ \bibnamefont
  {Campbell}},\ }\bibfield  {title} {\enquote {\bibinfo {title} {Quantum
  darwinism in a composite system: Objectivity versus classicality},}\ }\href
  {\doibase 10.3390/e23080995} {\bibfield  {journal} {\bibinfo  {journal}
  {Entropy}\ }\textbf {\bibinfo {volume} {23}} (\bibinfo {year} {2021}),\
  10.3390/e23080995}\BibitemShut {NoStop}%
\bibitem [{\citenamefont {Touil}\ \emph {et~al.}(2021)\citenamefont {Touil},
  \citenamefont {Yan}, \citenamefont {Girolami}, \citenamefont {Deffner},\ and\
  \citenamefont {Zurek}}]{touil2021eavesdropping}%
  \BibitemOpen
  \bibfield  {author} {\bibinfo {author} {\bibfnamefont {Akram}\ \bibnamefont
  {Touil}}, \bibinfo {author} {\bibfnamefont {Bin}\ \bibnamefont {Yan}},
  \bibinfo {author} {\bibfnamefont {Davide}\ \bibnamefont {Girolami}}, \bibinfo
  {author} {\bibfnamefont {Sebastian}\ \bibnamefont {Deffner}}, \ and\ \bibinfo
  {author} {\bibfnamefont {Wojciech~H.}\ \bibnamefont {Zurek}},\ }\bibfield
  {title} {\enquote {\bibinfo {title} {Eavesdropping on the decohering
  environment: Quantum darwinism, amplification, and the origin of objective
  classical reality},}\ }\href@noop {} {\  (\bibinfo {year} {2021})},\ \Eprint
  {http://arxiv.org/abs/2107.00035} {arXiv:2107.00035 [quant-ph]} \BibitemShut
  {NoStop}%
\bibitem [{\citenamefont {Ryan}\ \emph {et~al.}(2021)\citenamefont {Ryan},
  \citenamefont {Paternostro},\ and\ \citenamefont
  {Campbell}}]{RYAN2021127675}%
  \BibitemOpen
  \bibfield  {author} {\bibinfo {author} {\bibfnamefont {Eoghan}\ \bibnamefont
  {Ryan}}, \bibinfo {author} {\bibfnamefont {Mauro}\ \bibnamefont
  {Paternostro}}, \ and\ \bibinfo {author} {\bibfnamefont {Steve}\ \bibnamefont
  {Campbell}},\ }\bibfield  {title} {\enquote {\bibinfo {title} {Quantum
  darwinism in a structured spin environment},}\ }\href {\doibase
  https://doi.org/10.1016/j.physleta.2021.127675} {\bibfield  {journal}
  {\bibinfo  {journal} {Physics Letters A}\ }\textbf {\bibinfo {volume}
  {416}},\ \bibinfo {pages} {127675} (\bibinfo {year} {2021})}\BibitemShut
  {NoStop}%
\bibitem [{\citenamefont {Feynman}(1982)}]{Feynman1982}%
  \BibitemOpen
  \bibfield  {author} {\bibinfo {author} {\bibfnamefont {Richard~P.}\
  \bibnamefont {Feynman}},\ }\bibfield  {title} {\enquote {\bibinfo {title}
  {Simulating physics with computers},}\ }\href {\doibase 10.1007/BF02650179}
  {\bibfield  {journal} {\bibinfo  {journal} {International Journal of
  Theoretical Physics}\ }\textbf {\bibinfo {volume} {21}},\ \bibinfo {pages}
  {467--488} (\bibinfo {year} {1982})}\BibitemShut {NoStop}%
\bibitem [{\citenamefont {Lloyd}(1996)}]{doi:10.1126/science.273.5278.1073}%
  \BibitemOpen
  \bibfield  {author} {\bibinfo {author} {\bibfnamefont {Seth}\ \bibnamefont
  {Lloyd}},\ }\bibfield  {title} {\enquote {\bibinfo {title} {Universal quantum
  simulators},}\ }\href {\doibase 10.1126/science.273.5278.1073} {\bibfield
  {journal} {\bibinfo  {journal} {Science}\ }\textbf {\bibinfo {volume}
  {273}},\ \bibinfo {pages} {1073--1078} (\bibinfo {year} {1996})},\ \Eprint
  {http://arxiv.org/abs/https://www.science.org/doi/pdf/10.1126/science.273.5278.1073}
  {https://www.science.org/doi/pdf/10.1126/science.273.5278.1073} \BibitemShut
  {NoStop}%
\bibitem [{\citenamefont {Peruzzo}\ \emph {et~al.}(2014)\citenamefont
  {Peruzzo}, \citenamefont {McClean}, \citenamefont {Shadbolt}, \citenamefont
  {Yung}, \citenamefont {Zhou}, \citenamefont {Love}, \citenamefont
  {Aspuru-Guzik},\ and\ \citenamefont {O'Brien}}]{Peruzzo2014}%
  \BibitemOpen
  \bibfield  {author} {\bibinfo {author} {\bibfnamefont {Alberto}\ \bibnamefont
  {Peruzzo}}, \bibinfo {author} {\bibfnamefont {Jarrod}\ \bibnamefont
  {McClean}}, \bibinfo {author} {\bibfnamefont {Peter}\ \bibnamefont
  {Shadbolt}}, \bibinfo {author} {\bibfnamefont {Man-Hong}\ \bibnamefont
  {Yung}}, \bibinfo {author} {\bibfnamefont {Xiao-Qi}\ \bibnamefont {Zhou}},
  \bibinfo {author} {\bibfnamefont {Peter~J.}\ \bibnamefont {Love}}, \bibinfo
  {author} {\bibfnamefont {Al{\'a}n}\ \bibnamefont {Aspuru-Guzik}}, \ and\
  \bibinfo {author} {\bibfnamefont {Jeremy~L.}\ \bibnamefont {O'Brien}},\
  }\bibfield  {title} {\enquote {\bibinfo {title} {A variational eigenvalue
  solver on a photonic quantum processor},}\ }\href {\doibase
  10.1038/ncomms5213} {\bibfield  {journal} {\bibinfo  {journal} {Nature
  Communications}\ }\textbf {\bibinfo {volume} {5}},\ \bibinfo {pages} {4213}
  (\bibinfo {year} {2014})}\BibitemShut {NoStop}%
\bibitem [{\citenamefont {Kandala}\ \emph {et~al.}(2017)\citenamefont
  {Kandala}, \citenamefont {Mezzacapo}, \citenamefont {Temme}, \citenamefont
  {Takita}, \citenamefont {Brink}, \citenamefont {Chow},\ and\ \citenamefont
  {Gambetta}}]{Kandala2017}%
  \BibitemOpen
  \bibfield  {author} {\bibinfo {author} {\bibfnamefont {Abhinav}\ \bibnamefont
  {Kandala}}, \bibinfo {author} {\bibfnamefont {Antonio}\ \bibnamefont
  {Mezzacapo}}, \bibinfo {author} {\bibfnamefont {Kristan}\ \bibnamefont
  {Temme}}, \bibinfo {author} {\bibfnamefont {Maika}\ \bibnamefont {Takita}},
  \bibinfo {author} {\bibfnamefont {Markus}\ \bibnamefont {Brink}}, \bibinfo
  {author} {\bibfnamefont {Jerry~M.}\ \bibnamefont {Chow}}, \ and\ \bibinfo
  {author} {\bibfnamefont {Jay~M.}\ \bibnamefont {Gambetta}},\ }\bibfield
  {title} {\enquote {\bibinfo {title} {Hardware-efficient variational quantum
  eigensolver for small molecules and quantum magnets},}\ }\href {\doibase
  10.1038/nature23879} {\bibfield  {journal} {\bibinfo  {journal} {Nature}\
  }\textbf {\bibinfo {volume} {549}},\ \bibinfo {pages} {242--246} (\bibinfo
  {year} {2017})}\BibitemShut {NoStop}%
\bibitem [{\citenamefont {Garc{\'i}a-P{\'e}rez}\ \emph
  {et~al.}(2020)\citenamefont {Garc{\'i}a-P{\'e}rez}, \citenamefont {Rossi},\
  and\ \citenamefont {Maniscalco}}]{Garcia-Perez2020}%
  \BibitemOpen
  \bibfield  {author} {\bibinfo {author} {\bibfnamefont {Guillermo}\
  \bibnamefont {Garc{\'i}a-P{\'e}rez}}, \bibinfo {author} {\bibfnamefont
  {Matteo A.~C.}\ \bibnamefont {Rossi}}, \ and\ \bibinfo {author}
  {\bibfnamefont {Sabrina}\ \bibnamefont {Maniscalco}},\ }\bibfield  {title}
  {\enquote {\bibinfo {title} {Ibm q experience as a versatile experimental
  testbed for simulating open quantum systems},}\ }\href {\doibase
  10.1038/s41534-019-0235-y} {\bibfield  {journal} {\bibinfo  {journal} {npj
  Quantum Information}\ }\textbf {\bibinfo {volume} {6}},\ \bibinfo {pages} {1}
  (\bibinfo {year} {2020})}\BibitemShut {NoStop}%
\bibitem [{\citenamefont {Abbas}\ \emph {et~al.}(2020)\citenamefont {Abbas},
  \citenamefont {Sutter}, \citenamefont {Zoufal}, \citenamefont {Lucchi},
  \citenamefont {Figalli},\ and\ \citenamefont {Woerner}}]{abbas2020power}%
  \BibitemOpen
  \bibfield  {author} {\bibinfo {author} {\bibfnamefont {Amira}\ \bibnamefont
  {Abbas}}, \bibinfo {author} {\bibfnamefont {David}\ \bibnamefont {Sutter}},
  \bibinfo {author} {\bibfnamefont {Christa}\ \bibnamefont {Zoufal}}, \bibinfo
  {author} {\bibfnamefont {Aurélien}\ \bibnamefont {Lucchi}}, \bibinfo
  {author} {\bibfnamefont {Alessio}\ \bibnamefont {Figalli}}, \ and\ \bibinfo
  {author} {\bibfnamefont {Stefan}\ \bibnamefont {Woerner}},\ }\bibfield
  {title} {\enquote {\bibinfo {title} {The power of quantum neural networks},}\
  }\href@noop {} {\  (\bibinfo {year} {2020})},\ \Eprint
  {http://arxiv.org/abs/2011.00027} {arXiv:2011.00027 [quant-ph]} \BibitemShut
  {NoStop}%
\bibitem [{\citenamefont {Ciccarello}\ \emph {et~al.}(2021)\citenamefont
  {Ciccarello}, \citenamefont {Lorenzo}, \citenamefont {Giovannetti},\ and\
  \citenamefont {Palma}}]{ciccarello2021quantum}%
  \BibitemOpen
  \bibfield  {author} {\bibinfo {author} {\bibfnamefont {Francesco}\
  \bibnamefont {Ciccarello}}, \bibinfo {author} {\bibfnamefont {Salvatore}\
  \bibnamefont {Lorenzo}}, \bibinfo {author} {\bibfnamefont {Vittorio}\
  \bibnamefont {Giovannetti}}, \ and\ \bibinfo {author} {\bibfnamefont
  {G~Massimo}\ \bibnamefont {Palma}},\ }\bibfield  {title} {\enquote {\bibinfo
  {title} {Quantum collision models: open system dynamics from repeated
  interactions},}\ }\href@noop {} {\bibfield  {journal} {\bibinfo  {journal}
  {arXiv preprint arXiv:2106.11974}\ } (\bibinfo {year} {2021})}\BibitemShut
  {NoStop}%
\bibitem [{\citenamefont {Garc\'{\i}a-P\'erez}\ \emph
  {et~al.}(2020)\citenamefont {Garc\'{\i}a-P\'erez}, \citenamefont {Chisholm},
  \citenamefont {Rossi}, \citenamefont {Palma},\ and\ \citenamefont
  {Maniscalco}}]{PhysRevResearch.2.012061}%
  \BibitemOpen
  \bibfield  {author} {\bibinfo {author} {\bibfnamefont {Guillermo}\
  \bibnamefont {Garc\'{\i}a-P\'erez}}, \bibinfo {author} {\bibfnamefont
  {Dario~A.}\ \bibnamefont {Chisholm}}, \bibinfo {author} {\bibfnamefont
  {Matteo A.~C.}\ \bibnamefont {Rossi}}, \bibinfo {author} {\bibfnamefont
  {G.~Massimo}\ \bibnamefont {Palma}}, \ and\ \bibinfo {author} {\bibfnamefont
  {Sabrina}\ \bibnamefont {Maniscalco}},\ }\bibfield  {title} {\enquote
  {\bibinfo {title} {Decoherence without entanglement and quantum darwinism},}\
  }\href {\doibase 10.1103/PhysRevResearch.2.012061} {\bibfield  {journal}
  {\bibinfo  {journal} {Phys. Rev. Research}\ }\textbf {\bibinfo {volume}
  {2}},\ \bibinfo {pages} {012061} (\bibinfo {year} {2020})}\BibitemShut
  {NoStop}%
\bibitem [{\citenamefont {Chisholm}\ \emph {et~al.}(2021)\citenamefont
  {Chisholm}, \citenamefont {Garc{\'{\i}}a-P{\'{e}}rez}, \citenamefont {Rossi},
  \citenamefont {Palma},\ and\ \citenamefont {Maniscalco}}]{Chisholm_2021}%
  \BibitemOpen
  \bibfield  {author} {\bibinfo {author} {\bibfnamefont {Dario~A}\ \bibnamefont
  {Chisholm}}, \bibinfo {author} {\bibfnamefont {Guillermo}\ \bibnamefont
  {Garc{\'{\i}}a-P{\'{e}}rez}}, \bibinfo {author} {\bibfnamefont {Matteo A~C}\
  \bibnamefont {Rossi}}, \bibinfo {author} {\bibfnamefont {G~Massimo}\
  \bibnamefont {Palma}}, \ and\ \bibinfo {author} {\bibfnamefont {Sabrina}\
  \bibnamefont {Maniscalco}},\ }\bibfield  {title} {\enquote {\bibinfo {title}
  {Stochastic collision model approach to transport phenomena in quantum
  networks},}\ }\href {\doibase 10.1088/1367-2630/abd57d} {\bibfield  {journal}
  {\bibinfo  {journal} {New Journal of Physics}\ }\textbf {\bibinfo {volume}
  {23}},\ \bibinfo {pages} {033031} (\bibinfo {year} {2021})}\BibitemShut
  {NoStop}%
\bibitem [{\citenamefont {Vacchini}(2014)}]{doi:10.1142/S0219749914610115}%
  \BibitemOpen
  \bibfield  {author} {\bibinfo {author} {\bibfnamefont {Bassano}\ \bibnamefont
  {Vacchini}},\ }\bibfield  {title} {\enquote {\bibinfo {title} {General
  structure of quantum collisional models},}\ }\href {\doibase
  10.1142/S0219749914610115} {\bibfield  {journal} {\bibinfo  {journal}
  {International Journal of Quantum Information}\ }\textbf {\bibinfo {volume}
  {12}},\ \bibinfo {pages} {1461011} (\bibinfo {year} {2014})},\ \Eprint
  {http://arxiv.org/abs/https://doi.org/10.1142/S0219749914610115}
  {https://doi.org/10.1142/S0219749914610115} \BibitemShut {NoStop}%
\bibitem [{\citenamefont {Vacchini}(2016)}]{PhysRevLett.117.230401}%
  \BibitemOpen
  \bibfield  {author} {\bibinfo {author} {\bibfnamefont {Bassano}\ \bibnamefont
  {Vacchini}},\ }\bibfield  {title} {\enquote {\bibinfo {title} {Generalized
  master equations leading to completely positive dynamics},}\ }\href {\doibase
  10.1103/PhysRevLett.117.230401} {\bibfield  {journal} {\bibinfo  {journal}
  {Phys. Rev. Lett.}\ }\textbf {\bibinfo {volume} {117}},\ \bibinfo {pages}
  {230401} (\bibinfo {year} {2016})}\BibitemShut {NoStop}%
\bibitem [{\citenamefont {Vacchini}(2020)}]{Vacchini2020}%
  \BibitemOpen
  \bibfield  {author} {\bibinfo {author} {\bibfnamefont {Bassano}\ \bibnamefont
  {Vacchini}},\ }\bibfield  {title} {\enquote {\bibinfo {title} {Quantum
  renewal processes},}\ }\href {\doibase 10.1038/s41598-020-62260-z} {\bibfield
   {journal} {\bibinfo  {journal} {Scientific Reports}\ }\textbf {\bibinfo
  {volume} {10}},\ \bibinfo {pages} {5592} (\bibinfo {year}
  {2020})}\BibitemShut {NoStop}%
\bibitem [{\citenamefont {Lindblad}(1976)}]{Lindblad1976}%
  \BibitemOpen
  \bibfield  {author} {\bibinfo {author} {\bibfnamefont {G.}~\bibnamefont
  {Lindblad}},\ }\bibfield  {title} {\enquote {\bibinfo {title} {On the
  generators of quantum dynamical semigroups},}\ }\href {\doibase
  10.1007/BF01608499} {\bibfield  {journal} {\bibinfo  {journal}
  {Communications in Mathematical Physics}\ }\textbf {\bibinfo {volume} {48}},\
  \bibinfo {pages} {119--130} (\bibinfo {year} {1976})}\BibitemShut {NoStop}%
\bibitem [{\citenamefont {Gorini}\ \emph {et~al.}(1976)\citenamefont {Gorini},
  \citenamefont {Kossakowski},\ and\ \citenamefont
  {Sudarshan}}]{doi:10.1063/1.522979}%
  \BibitemOpen
  \bibfield  {author} {\bibinfo {author} {\bibfnamefont {Vittorio}\
  \bibnamefont {Gorini}}, \bibinfo {author} {\bibfnamefont {Andrzej}\
  \bibnamefont {Kossakowski}}, \ and\ \bibinfo {author} {\bibfnamefont
  {E.~C.~G.}\ \bibnamefont {Sudarshan}},\ }\bibfield  {title} {\enquote
  {\bibinfo {title} {Completely positive dynamical semigroups of n‐level
  systems},}\ }\href {\doibase 10.1063/1.522979} {\bibfield  {journal}
  {\bibinfo  {journal} {Journal of Mathematical Physics}\ }\textbf {\bibinfo
  {volume} {17}},\ \bibinfo {pages} {821--825} (\bibinfo {year} {1976})},\
  \Eprint
  {http://arxiv.org/abs/https://aip.scitation.org/doi/pdf/10.1063/1.522979}
  {https://aip.scitation.org/doi/pdf/10.1063/1.522979} \BibitemShut {NoStop}%
\bibitem [{\citenamefont {Breuer}\ \emph {et~al.}(2009)\citenamefont {Breuer},
  \citenamefont {Laine},\ and\ \citenamefont {Piilo}}]{PhysRevLett.103.210401}%
  \BibitemOpen
  \bibfield  {author} {\bibinfo {author} {\bibfnamefont {Heinz-Peter}\
  \bibnamefont {Breuer}}, \bibinfo {author} {\bibfnamefont {Elsi-Mari}\
  \bibnamefont {Laine}}, \ and\ \bibinfo {author} {\bibfnamefont {Jyrki}\
  \bibnamefont {Piilo}},\ }\bibfield  {title} {\enquote {\bibinfo {title}
  {Measure for the degree of non-markovian behavior of quantum processes in
  open systems},}\ }\href {\doibase 10.1103/PhysRevLett.103.210401} {\bibfield
  {journal} {\bibinfo  {journal} {Phys. Rev. Lett.}\ }\textbf {\bibinfo
  {volume} {103}},\ \bibinfo {pages} {210401} (\bibinfo {year}
  {2009})}\BibitemShut {NoStop}%
\bibitem [{\citenamefont {\ifmmode \check{R}\else
  \v{R}\fi{}eh\'a\ifmmode~\check{c}\else \v{c}\fi{}ek}\ \emph
  {et~al.}(2007)\citenamefont {\ifmmode \check{R}\else
  \v{R}\fi{}eh\'a\ifmmode~\check{c}\else \v{c}\fi{}ek}, \citenamefont {Hradil},
  \citenamefont {Knill},\ and\ \citenamefont {Lvovsky}}]{PhysRevA.75.042108}%
  \BibitemOpen
  \bibfield  {author} {\bibinfo {author} {\bibfnamefont {Jaroslav}\
  \bibnamefont {\ifmmode \check{R}\else \v{R}\fi{}eh\'a\ifmmode~\check{c}\else
  \v{c}\fi{}ek}}, \bibinfo {author} {\bibfnamefont {Zden\ifmmode
  \check{e}\else~\v{e}\fi{}k}\ \bibnamefont {Hradil}}, \bibinfo {author}
  {\bibfnamefont {E.}~\bibnamefont {Knill}}, \ and\ \bibinfo {author}
  {\bibfnamefont {A.~I.}\ \bibnamefont {Lvovsky}},\ }\bibfield  {title}
  {\enquote {\bibinfo {title} {Diluted maximum-likelihood algorithm for quantum
  tomography},}\ }\href {\doibase 10.1103/PhysRevA.75.042108} {\bibfield
  {journal} {\bibinfo  {journal} {Phys. Rev. A}\ }\textbf {\bibinfo {volume}
  {75}},\ \bibinfo {pages} {042108} (\bibinfo {year} {2007})}\BibitemShut
  {NoStop}%
\bibitem [{\citenamefont {Holevo}(1973)}]{holevo1973bounds}%
  \BibitemOpen
  \bibfield  {author} {\bibinfo {author} {\bibfnamefont {Alexander~Semenovich}\
  \bibnamefont {Holevo}},\ }\bibfield  {title} {\enquote {\bibinfo {title}
  {Bounds for the quantity of information transmitted by a quantum
  communication channel},}\ }\href@noop {} {\bibfield  {journal} {\bibinfo
  {journal} {Problemy Peredachi Informatsii}\ }\textbf {\bibinfo {volume}
  {9}},\ \bibinfo {pages} {3--11} (\bibinfo {year} {1973})}\BibitemShut
  {NoStop}%
\bibitem [{\citenamefont {Korbicz}\ \emph {et~al.}(2014)\citenamefont
  {Korbicz}, \citenamefont {Horodecki},\ and\ \citenamefont
  {Horodecki}}]{PhysRevLett.112.120402}%
  \BibitemOpen
  \bibfield  {author} {\bibinfo {author} {\bibfnamefont {J.~K.}\ \bibnamefont
  {Korbicz}}, \bibinfo {author} {\bibfnamefont {P.}~\bibnamefont {Horodecki}},
  \ and\ \bibinfo {author} {\bibfnamefont {R.}~\bibnamefont {Horodecki}},\
  }\bibfield  {title} {\enquote {\bibinfo {title} {Objectivity in a noisy
  photonic environment through quantum state information broadcasting},}\
  }\href {\doibase 10.1103/PhysRevLett.112.120402} {\bibfield  {journal}
  {\bibinfo  {journal} {Phys. Rev. Lett.}\ }\textbf {\bibinfo {volume} {112}},\
  \bibinfo {pages} {120402} (\bibinfo {year} {2014})}\BibitemShut {NoStop}%
\bibitem [{\citenamefont {Horodecki}\ \emph {et~al.}(2015)\citenamefont
  {Horodecki}, \citenamefont {Korbicz},\ and\ \citenamefont
  {Horodecki}}]{PhysRevA.91.032122}%
  \BibitemOpen
  \bibfield  {author} {\bibinfo {author} {\bibfnamefont {R.}~\bibnamefont
  {Horodecki}}, \bibinfo {author} {\bibfnamefont {J.~K.}\ \bibnamefont
  {Korbicz}}, \ and\ \bibinfo {author} {\bibfnamefont {P.}~\bibnamefont
  {Horodecki}},\ }\bibfield  {title} {\enquote {\bibinfo {title} {Quantum
  origins of objectivity},}\ }\href {\doibase 10.1103/PhysRevA.91.032122}
  {\bibfield  {journal} {\bibinfo  {journal} {Phys. Rev. A}\ }\textbf {\bibinfo
  {volume} {91}},\ \bibinfo {pages} {032122} (\bibinfo {year}
  {2015})}\BibitemShut {NoStop}%
\bibitem [{\citenamefont {Le}\ and\ \citenamefont
  {Olaya-Castro}(2019)}]{PhysRevLett.122.010403}%
  \BibitemOpen
  \bibfield  {author} {\bibinfo {author} {\bibfnamefont {Thao~P.}\ \bibnamefont
  {Le}}\ and\ \bibinfo {author} {\bibfnamefont {Alexandra}\ \bibnamefont
  {Olaya-Castro}},\ }\bibfield  {title} {\enquote {\bibinfo {title} {Strong
  quantum darwinism and strong independence are equivalent to spectrum
  broadcast structure},}\ }\href {\doibase 10.1103/PhysRevLett.122.010403}
  {\bibfield  {journal} {\bibinfo  {journal} {Phys. Rev. Lett.}\ }\textbf
  {\bibinfo {volume} {122}},\ \bibinfo {pages} {010403} (\bibinfo {year}
  {2019})}\BibitemShut {NoStop}%
\end{thebibliography}%

\end{document}